\documentclass[12pt,a4paper]{article}
\usepackage{amssymb}
\usepackage{epsf}
\begin{document}

\title{Bubble nucleation and growth in very strong
cosmological phase transitions}
\author{\large  Ariel M\'{e}gevand\thanks{%
Member of CONICET, Argentina. E-mail address: megevand@mdp.edu.ar}~ and
Santiago Ram\'{\i}rez \\[0.5cm]
\normalsize \it IFIMAR (UNMdP-CONICET)\\ \normalsize \it Departamento de
F\'{\i}sica, Facultad de Ciencias Exactas y Naturales, \\ \normalsize \it
UNMdP, De\'{a}n Funes 3350, (7600) Mar del Plata, Argentina }
\date{}
\maketitle

\begin{abstract}
Strongly first-order phase transitions, i.e., those with a large order
parameter, are characterized by a considerable supercooling and high
velocities of phase transition fronts. A very strong phase transition may
have important cosmological consequences due to the departures from
equilibrium caused in the plasma. In general, there is a limit to the
strength, since the metastability of the old phase may prevent the
transition to complete. Near this limit, the bubble nucleation rate
achieves a maximum and thus departs from the widely assumed behavior in
which it grows exponentially with time. We study the dynamics of this kind
of phase transitions. We show that in some cases a gaussian approximation
for the nucleation rate is more suitable, and in such a case we solve
analytically the evolution of the phase transition. We compare the
gaussian and exponential approximations with realistic cases and we
determine their ranges of validity. We also discuss the implications for
cosmic remnants such as gravitational waves.
\end{abstract}

\section{Introduction}

A cosmological first-order phase transition may have several observable
consequences, such as the generation of the baryon asymmetry of the universe
(for a review, see \cite{ewbaryo}), the formation of topological defects
\cite{vs94}, or the production of gravitational waves (see \cite{maggiore}
for a review). The estimation of these relics involves computing the
development of the phase transition in order to determine quantities such as
the bubble wall velocity, the number density of nucleated bubbles, and the
distribution of bubble sizes. Depending on the computational method, several
approximations are usually necessary. In particular, it is customary to
consider a constant wall velocity and specific analytical forms for the
nucleation rate as a function of time. Among the latter, the most common
assumptions are an exponentially growing rate, a constant rate, or just a
simultaneous nucleation. The validity of these approximations depends on the
strength of the phase transition.

For a first-order phase transition, the free energy density $\mathcal{F}$ is
a function of an order parameter $\phi $ and has two minima separated by a
barrier. The stable phase corresponds to the absolute minimum, while the
metastable phase corresponds to a local minimum. For concreteness, we shall
consider the case in which at high temperature $T$ we have a stable minimum
$\phi _{+}=0$, corresponding to a symmetric phase. At low temperatures, this
minimum becomes metastable, while a broken-symmetry minimum $\phi _{-}$
becomes the stable one (we shall use the subscripts $\pm$ for quantities
corresponding to each of these phases). At the critical temperature $T_{c}$
the two minima of $\mathcal{F}(\phi,T)$ are degenerate. Several quantities,
such as the energy density, are different in each phase. Therefore, if we
assume that the phase transition takes place at $T=T_{c}$, these quantities
are discontinuous functions of $T$. The discontinuities depend on the jump
of $\phi $. Thus, the value of $\phi _{-}/T$ at $T=T_{c}$ can be regarded as
an order parameter. The phase transition is usually said to be weakly first
order if $\phi _{-}/T\ll 1$ and strongly first order if $\phi _{-}/T\gtrsim
1$. We shall be mainly interested in the case of very strong phase
transitions, for which $\phi _{-}/T\gg 1$.

At $T= T_{c}$ or higher, the bubble nucleation rate per unit volume, $\Gamma
$, vanishes. Therefore, the phase transition does not  occur exactly at
$T=T_{c}$. Below $T_{c}$, the nucleation rate grows continuously from
$\Gamma=0$, and may reach, in principle, a value $\Gamma \sim T_c^4\sim
v^4$, where $v$ is the energy scale characterizing the phase transition. In
many cases there is a temperature $T_{0}<T_{c}$ at which the barrier
separating the two minima disappears and the local minimum $\phi _{+}=0$
becomes a maximum (see \cite{quiros} for a review). At this temperature, the
phase transition would proceed through spinodal decomposition rather than by
bubble nucleation. Nevertheless, before reaching $T=T_{0}$ we would have
$\Gamma \sim v^{4}$, which is extremely large. Indeed, the development of
the phase transition depends crucially on the relation between $\Gamma $ and
the expansion rate $H$ \cite{tww92}, since the number of bubbles $N$
nucleated in a causal volume $V\sim H^{-3}$ in a cosmological time $t\sim
H^{-1}$ will be $N\sim \Gamma H^{-4}$. While $\Gamma$ varies significantly,
$H$ is roughly given by $H\sim T^2/M_P\sim v^{2}/M_{P}$, where $M_{p}$ is
the Planck mass. Thus, for $T$ close enough to $T_{c}$ we will have $ \Gamma
\ll H^{4}$ and the nucleation of bubbles will be too slow, while for
$\Gamma\sim v^4$ we have $\Gamma H^{-4}\sim (M_{P}/v)^{4}\gg 1$, since in
most cases $v$ is several orders of magnitude below $M_{P}$. Therefore, the
phase transition will generally occur at an intermediate temperature between
$T_c$ and $T_0$, such that $\Gamma H^{-4}\sim 1$, and, due to the rapid
growth of $\Gamma $, it will end in a time $\Delta t\ll t$.

For a weakly first-order phase transition, the barrier between minima is
relatively small and the temperature $T_{0}$ is very close to $T_{c}$. On
the other hand, for stronger phase transitions we have a larger barrier
which persists at smaller temperatures or even at $T=0$. In such a case
there is no temperature $T_{0}$ (see \cite{ewur} for examples of such
models), and the phase transition can last longer.

Since in many cases the duration of the phase transition is short, the
nucleation rate is sometimes approximated by a constant. This is convenient,
e.g., for involved computations such as numerical simulations, in which
sometimes it is even necessary to nucleate all the bubbles simultaneously.
However, it is the variation of $\Gamma $ with time what actually determines
the dynamics of the transition. In general, the nucleation rate is of the
form $\Gamma =A\exp (-S) $, where the quantities $S$ and $A$ are functions
of temperature and, hence, of time. Nevertheless, the variation of $\Gamma$
is dominated by the exponent $S$. A common approximation is to assume that
the duration of the phase transition is short enough to justify the
expansion of $S$ to linear order about a suitable time $t_{\ast }$. This
gives a nucleation rate of the form
\begin{equation}
\Gamma =\Gamma _{\ast }\exp [\beta (t-t_{\ast })],  \label{gammaexplin}
\end{equation}
where $\Gamma _{\ast }=\Gamma (t_{\ast })$ and
\begin{equation}
\beta =-(dS/dt)|_{t_{\ast }}=H(TdS/dT)|_{t_{\ast }}.  \label{beta}
\end{equation}
In the last equality, the temperature variation $dT/T=-da/a$ is
assumed\footnote{More generally, from the equation for the entropy density,
$ds/dt=-3sH$, we obtain $\beta /H=(dS/dT)_{\ast }(3sdT/ds)_{\ast }$. For
$s\propto T^{3}$ the last factor yields $T$, and we obtain
Eq.~(\ref{beta}).}, where $a$ is the scale factor. The quantity $\beta $ is
usually positive. This is because, as the temperature decreases from $T_c$
to $T_0$, the barrier between minima decreases until disappearing. Hence,
the nucleation rate monotonically grows with time and the exponent $S$
decreases. With the approximation (\ref{gammaexplin}), the parameter $\beta
$ determines the time scale for the phase transition. Thus, the duration is
given by $ \Delta t\sim \beta ^{-1}$ and the typical bubble size is given by
$R_{b}\sim v_{w}\beta ^{-1}$, where $v_{w}$ is the bubble wall velocity. The
latter can be assumed to be a constant if the duration of the phase
transition is short enough.

The expression (\ref{gammaexplin}) is widely used as an approximation for
the nucleation rate, either to simplify numerical computations or to obtain
analytic results. One might expect that the essential conditions on which it
is based, namely, that $S(t)$ can be linearized and that $\beta$ is
positive, are quite generally valid. However, both of these conditions may
break down. One case in which this happens is that of a very strong phase
transition, such that the barrier persists at $T=0$. In such a case, as $T$
descends from $T_{c}$ the nucleation rate will initially grow (as the
barrier decreases), but for low enough temperatures $\Gamma $ will stop
growing and will begin to decrease, since the presence of a barrier at small
temperatures will prevent the nucleation of bubbles. Thus, in this case
$\Gamma $ has a maximum and $S$ has a minimum at a certain temperature
$T_{m}$. Assuming that the system reaches such amounts of supercooling, the
linear approximation for $S(t)$ will not be valid at the corresponding time
$t=t_{m}$. In such a case, the sign of $\beta $ will be different on each
side of $t_{m}$, and $\Gamma $ will clearly depart from the form
(\ref{gammaexplin}).

This approximation may also break down due to reheating. Indeed, the energy
density difference between the two phases (latent heat) is released at the
bubble walls and reheats the surrounding plasma. Such a temperature increase
may cause a decrease of $\Gamma $. Nevertheless, the effect will depend on
the wall velocity. Indeed, if the phase transition fronts propagate as
detonations with supersonic velocities, the fluid in front of them remains
unperturbed. Thus, the reheating occurs only inside the bubbles, while in
the metastable phase the nucleation rate continues growing due to the
adiabatic cooling. The same holds for a runaway wall, i.e., a wall which has
not reached a steady state and propagates at almost the speed of light. For
smaller wall velocities the bubbles expand as deflagrations. A deflagration
wall is preceded by a shock wave which carries the released energy outside
the bubble. In this case, the plasma in the metastable phase will be
reheated, and the nucleation rate will be suppressed. This is a complex
scenario, since the reheating is inhomogeneous, causing an inhomogeneous
$\Gamma $. Besides, the growth of a bubble will be affected by the latent
heat released by other bubbles, which complicates further the dynamics. We
shall address this case elsewhere and focus here on very strong phase
transitions, for which we generally have detonations or runaway walls.

We shall study the dynamics of such very strong transitions. The main aim of
this paper is to assess the limit of validity of the approximation
(\ref{gammaexplin}), and to give an alternative analytic description of the
transition, based on a simple approximation for the nucleation rate beyond
this limit. The plan of the paper is the following. In the next section we
identify the main aspects of the dynamics which are relevant for the
possible cosmic remnants. In Sec.~\ref{dyn} we use a realistic toy model to
study the dynamics of the nucleation and growth of bubbles in very strong
phase transitions. In Sec.~\ref{anap} we consider analytic approximations
for the nucleation rate. We show that, in the very strong limit, the
exponential approximation breaks down and a gaussian approximation is more
appropriate. The latter allows to obtain analytical results for the dynamics
as much as the former. We analyse the ranges of validity of these
approximations. In section \ref{aplic} we compare the analytic results with
the numerical computation, and we discuss the effect of using different
approximations on the formation of cosmic remnants. Our conclusions are
summarized in Sec.~\ref{conclu}.

\section{Cosmological consequences of a very strong phase transition}
\label{conseq}

The most important consequence of a very strong phase transition is probably
the generation of sizeable gravitational radiation \cite{tw90}.
Gravitational waves (GWs) are produced by the collisions of bubble walls
\cite{ktw92} as well as by the motions caused in the plasma \cite{kkt94}
(see also \cite{dgn02,gkk07,cds09}). The process of bubble collisions is
usually computed with the envelope approximation \cite{kt93}, in which
bubbles are modeled by thin spherical shells which expand with a constant
velocity. This approximation has been used in numerical simulations
\cite{kt93,hk08,w16} in which the bubbles are nucleated at random positions
with a rate given by Eq.~(\ref{gammaexplin}). More recently, an analytic
approach using the same approximations was considered \cite{jt16}. The
general result is a GW spectrum which depends on the wall velocity and the
constant $\beta$. The quantity to be computed is the normalized energy
density per logarithmic frequency of GWs,
$\Omega_{GW}(f)=(1/\rho_{\mathrm{tot}})d\rho_{GW}/d(\log f)$, where
$\rho_{\mathrm{tot}}$ is the total energy density in the plasma. For
$v_w\simeq 1$, the peak frequency and amplitude from the envelope
approximation are \cite{hk08}
\begin{equation}
\tilde{f}_{\mathrm{coll} }\simeq 0.23\,\beta , \;\;
 \tilde\Omega _{\mathrm{coll}}\simeq 0.08\left( \rho_{\mathrm{wall}}
 /{\rho _{\mathrm{tot}}}\right) ^{2}\left( {H}/{\beta }\right)^{2}.
\label{GWcol}
\end{equation}
where $\rho_{\mathrm{wall}}$ is the average energy density which accumulates
in the thin shells. This corresponds to the gradient energy of the Higgs
field, which is a free parameter in the simulation. Although fluid motions
generally produce a larger signal of GWs, bubble collisions become important
for runaway walls, since in this case a significant fraction of the energy
goes to the wall rather than to the fluid.

The effects of the fluid motions could in principle be calculated using the
envelope approximation, if one assumes that the energy is concentrated in a
thin region \cite{kkt94}. However, this is generally not the case (for a
recent discussion, see \cite{w16}). Moreover, the fluid motions remain after
the bubbles have filled all the space and their walls have already
disappeared. Lattice simulations \cite{hhrw} of the field-fluid system
suggest that the strongest source of GWs are the sound waves generated in
the plasma. A fit to these results is given in Ref.~\cite{elisasci}. For the
peak of the spectrum we have
\begin{equation}
\tilde{f}_{sw}\simeq 3.4/d,
\;\;
\tilde{\Omega}_{sw}\simeq 0.02\left({\rho _{\mathrm{fl}}}
/{\rho _{\mathrm{tot}}}\right)^{2}{Hd},
\label{GWsw}
\end{equation}
where $\rho_{\mathrm{fl}}$ is the energy density in bulk fluid motions, and
$d$ is the average distance between bubble centers. It is important to
mention that in Ref.~\cite{elisasci} $\tilde f$ and $\tilde \Omega$ are
actually given in terms of $\beta$ rather than $d$, although in the
simulations all the bubbles were nucleated simultaneously and thus the scale
is set by the bubble separation. The dependence on $\beta $ was imposed by
the replacement $d=(8\pi )^{1/3}v_{w}/\beta$, which corresponds to the
analytic approximation we discuss in Sec.~\ref{anap}.

Another possible remnant of a cosmological phase transition is a network of
cosmic strings or other topological defects. Moreover, these defects can
only be formed in a phase transition of the universe. In the simplest case
in which a global $U(1)$ symmetry is spontaneously broken, the phase angle
$\theta$ of the Higgs field takes different values in uncorrelated regions.
For a first-order phase transition, these regions are the nucleated bubbles.
As bubbles meet and the variation of the phase from one domain to another is
smoothed out, a vortex or string may be trapped in the meeting point of
several bubbles \cite{k76}. This determines a segment of string whose length
depends on the bubble size. Therefore, the strings will have the shape of
random walks, where the step size depends on the size distribution of
bubbles at the end of the phase transition.

The statistical properties of the system of strings were studied using Monte
Carlo simulations, in which a large volume is divided in cubic cells of
fixed size, corresponding to the correlation length $\xi$ \cite{vv84}. The
result is that about 20\% of the total length of strings is contributed by
closed loops, while the rest are infinite strings. The  number density of
loops per loop length is given by $dn/dl\sim \xi^{-3/2}l^{-5/2}$. For a
first-order phase transition, the characteristic length $\xi$ is roughly
given by the average distance $d$ between bubble centers. More realistic
simulations of the phase transition were performed in 2+1 dimensions,
nucleating bubbles at random positions and at random times
\cite{bkvv95,f98}. The resulting number of defects  per unit volume depends
on the number density of bubbles and the wall velocity. Numerical
simulations for gauge symmetry braking were also performed (see e.g.
\cite{bov07}). The probability for trapping a defect actually depends on the
dynamics of the Higgs phase $\theta$. This dynamics is important for slow
bubble walls, since the phase difference between two bubbles could
equilibrate before a third bubble meets them. This results in the formation
of a smaller number of defects for smaller velocities. On the other hand,
for very strong phase transitions we generally have $v_w\simeq 1$, in which
case the dynamics of phase equilibration can be neglected (see, e.g.,
\cite{lf01}).

Other cosmological consequences of the phase transition depend on similar
aspects of the dynamics. For instance, the formation of the baryon asymmetry
of the universe in the electroweak phase transition (electroweak
baryogenesis) depends strongly on the velocity of bubble walls. Furthermore,
in this case, variations of the wall velocity may lead to baryon
inhomogeneities \cite{h95,ma05}. The shape of these inhomogeneities will be
similar inside each bubble, and their size will be determined by the bubble
radius. We shall not discuss this possibility here, since the mechanism of
electroweak baryogenesis generally requires small velocities, $v_w\sim
10^{-2}$-$10^{-1}$ \cite{ck00}, which are not likely in the case of very
strong phase transitions (see, however, \cite{n11,cn12,mm14,v16}).  In any
case, all these examples show that two aspects of the phase transition
dynamics are relevant for the formation of cosmic remnants, namely, the wall
velocity and the distribution of bubble sizes.

A complete calculation of the remnants involves the whole evolution of the
phase transition. Notice that, in order to take into account the global
dynamics (i.e., the expansion and collisions of many bubbles) in a realistic
way, the dynamics of bubble nucleation is generally simplified. This is
accomplished by considering specific forms for the nucleation rate, which
are not necessarily realistic. Thus, for lattice simulations it is usual to
either nucleate all the bubbles simultaneously or at a constant rate, while
in less time-consuming simulations the exponential approximation
(\ref{gammaexplin}) is often used.

\section{Phase transition dynamics}  \label{dyn}

In order to investigate the general features of the phase transition
dynamics, we shall consider a simple free energy depending on a few
parameters. For several considerations it actually suffices to use simple
functions $\mathcal{F}_+(T)$ and $\mathcal{F}_-(T)$ to describe the free
energy density of each phase, which in a real model are given by the
corresponding minima, $\mathcal{F}_{\pm }(T)=\mathcal{F}(\phi _{\pm },T)$.
The pressure is given by $p_{\pm }=-\mathcal{F}_{\pm }$ and the energy
density by $\rho _{\pm }=Ts_{\pm }-p_{\pm }$, with $s_{\pm}=dp_{\pm}/dT$.
Thus, these functions define the equation of state (EOS) in each phase.
However, a realistic treatment of bubble nucleation requires specifying the
whole dependence of $\mathcal{F}$ on the order-parameter field $\phi $.
Therefore, we will consider a free energy density of the form
\begin{equation}
\mathcal{F}(\phi ,T)=V(\phi ,T)-\frac{\pi ^{2}}{90}g_{\ast }T^{4}+\rho _{V},
\label{freeen}
\end{equation}
where the first term is the field-dependent part, which vanishes for
$\phi=0$, the second term corresponds to massless particles which do not
interact with $\phi $ (with a total number  of effective degrees of freedom
$g_{\ast }$), and the constant $\rho_V$ is the vacuum energy density for
$\phi=0$. Therefore, the total energy density of the phase $+$ is given by
$\rho _{+}=\rho _{V}+\rho _{R} $, where
\begin{equation}
\rho _{R}=\frac{\pi ^{2}}{30}g_{\ast }T^{4}\label{rhoR}
\end{equation}
is the radiation energy density. The Hubble rate will be essentially
determined by the quantity $\rho_+$, while the nucleation rate will depend
only on the $\phi $-dependent part $V(\phi,T)$.

\subsection{Effective potential  and nucleation rate}

The simplest physical model which may provide very strong phase transitions
is given by a finite-temperature effective potential of the form
\begin{equation}
V(\phi ,T)=D(T^{2}-T_{0}^{2})\phi ^{2}-(ET+A)\phi ^{3}+\frac{\lambda }{4}
\phi ^{4}.  \label{potef}
\end{equation}
For $A=0$, Eq.~(\ref{potef}) has the well-known form of the high-temperature
expansion of the one-loop effective potential, for a  tree-level potential
of the form $V_0=-m^2\phi^2+\lambda_0\phi^4$ (see, e.g.,
\cite{ah92,dlhll92}). It is well known that tree-level modifications of this
potential can easily give strong phase transitions, and a simple example is
the cubic term $-A\phi ^{3}$. The high-temperature minimum of $V$ is
$\phi_+=0$, the low-temperature minimum is given by
\begin{equation}
\phi _{-}(T)=\frac{3(ET+A)}{2\lambda }\left[ 1+\sqrt{1-\frac{8\lambda
D(T^{2}-T_{0}^{2})}{9(ET+A)^{2}}}\right] ,  \label{fime}
\end{equation}
and the critical temperature is given by
\begin{equation}
T_{c}=\frac{1}{\lambda D-E^{2}}\left[ AE+\sqrt{\lambda D}\sqrt{A^{2}
+(\lambda D-E^{2})T_{0}^{2}}\right] .
\end{equation}
The strength of the phase transition is usually measured by the quantity
$\phi_c/T_c$, with $\phi_c=\phi _{-}(T_{c})$. We have\footnote{Another
useful relation is $\phi _{c}^{2}=(4D/\lambda)(T_{c}^{2}-T_{0}^{2})$.}
\begin{equation}
\frac{\phi _{c}}{T_{c}}=\frac{2(E+A/T_{c})}{\lambda }.  \label{fic}
\end{equation}
A further quantity of interest will be the latent heat $\rho
_{+}(T_{c})-\rho _{-}(T_{c})=T\frac{\partial V}{\partial T}(\phi
_{c},T_{c})$. We have
\begin{equation}
\frac{L}{T_{c}^{4}}=2D\left( \frac{\phi _{c}}{T_{c}}\right) ^{2}-E\left(
\frac{\phi _{c}}{T_{c}}\right) ^{3}.  \label{lh}
\end{equation}

The strength of the phase transition is directly related to the cubic term
of the potential. Indeed, for $A=0$ and $E=0$ we have $T_c=T_0$, $\phi_c=0$,
and  the phase transition is second order. In this case, for $T<T_{c}$ the
minimum moves continuously to $\phi _{-}(T)>0$, while the point $\phi =0$
becomes a maximum. For $A>0$ or $E>0$ the phase transition is first order.
In this case, at $T<T_{0}$ the point $ \phi =0$ is  a maximum and the only
minimum is $\phi_-$. At $T>T_{0}$ the maximum becomes a minimum, and the two
minima coexist up to a temperature $T_{1}$, where  the minimum $\phi _{-}$
disappears. The critical temperature $T_{c}$ lies between $T_{0}$ and
$T_{1}$. As can be seen in Eq.~(\ref{fic}), for $A=0$ the quantity
$\phi_c/T_c$ is proportional to the parameter $E$. This means that in this
case the first-order nature of the phase transition is due to a
fluctuation-induced cubic term, which is not present in the tree-level
potential. In a realistic model, $E$ depends cubically on the couplings of
bosons with $\phi $, and it is difficult to obtain a relatively strong phase
transition (i.e., one in which $\phi _{c}/T_{c}> 1$). In contrast,
Eq.~(\ref{fic}) shows that adding a tree-level cubic term increases the
order parameter, and we obtain a strongly first-order phase transition for
$A/T_c\sim\lambda$. Furthermore, runaway walls are only possible if the
first-order nature of the phase transition is due to zero-temperature terms
\cite{bm09}.

It is convenient to relate the parameters of the potential  to physical
quantities. In the first place, the zero-temperature minimum $v=\phi_-(T=0)$
is given as a function of the parameters by Eq.~(\ref{fime}). If we want to
vary the shape of the potential without changing the characteristic scale of
the theory, then we must fix the value of $v$. This sets the value of the
spinodal-decomposition temperature,
\begin{equation}
T_{0}^{2}=\frac{\lambda -3A/v}{2D}v^{2}.  \label{T0}
\end{equation}
This temperature decreases for increasing $A/v$, indicating that the barrier
between minima persists at lower temperatures. Notice that for $A/v>\lambda
/3$ we have $T_{0}^{2}<0$, which indicates that in this case the potential
has a barrier even at $T=0$. Hence, large values of $A/v$ will imply large
amounts of supercooling. On the other hand, the value of $\lambda $ should
be such that the curvature of the potential is naturally given by the scale
$v$. Thus, we may choose as a physical parameter the scalar mass
\begin{equation}
m_{\phi }^{2}=\frac{\partial ^{2}V}{\partial \phi ^{2}}
(v,0)=2\lambda v^{2}-3Av,
\end{equation}
and set  $m_{\phi }\sim v$. The value of the dimensionless parameter $E$, as
well as the ratio $A/v$ may be determined by choosing a definite value for
$\phi _{c}/T_c$ (e.g., demanding a strongly first-order phase transition).
The value of $A/v$ has a stronger effect on the dynamics, as it affects also
the spinodal temperature $T_0$, which vanishes for $A/v=(m_{\phi}/v)^{2}/3$.
Finally, the parameter $D$ can be set as a function the latent heat using
Eq.~(\ref{lh}).

The constant $ \rho _{V}$ in Eq.~(\ref{freeen}) is determined by the
condition that the vacuum energy vanishes in the true vacuum at $T=0$, i.e.,
$\rho_V=-V(v,0)$. This gives
\begin{equation}
\rho _{V}=\frac{1}{4}(\lambda -2A/v)v^{4}.
\end{equation}
We see that for $A/v>\lambda /2$ the false vacuum energy becomes negative.
This actually means that for large $A/v$ the minimum $\phi =0$ becomes
stable even at zero temperature, while the minimum $v$ becomes metastable.
Indeed, for $A/v=\lambda /2$ the critical temperature vanishes. This sets an
upper limit  $A/v=(m_{\phi }/v)^{2} $.

The probability of bubble nucleation per unit volume per unit time is of the
form \cite{affleck}
\begin{equation}
\Gamma (T)=A(T)\,e^{-S(T)},  \label{gamma}
\end{equation}
where $S(T)$ is an instanton action and $A(T)$ is a dimensional determinant.
In the limit of very low temperatures the nucleation rate can be
approximated by the false-vacuum decay (quantum tunneling) result, in which
$S$ is the four-dimensional instanton action $S_4$ and the prefactor is
given by $A\simeq v^{4}[S_{4}/(2\pi )]^{2}$ \cite{coleman}. The action is
given by
\begin{equation}
S_{4}=2\pi ^{2}\int_{0}^{\infty }r^{3}dr\left[ \frac{1}{2}\left( \frac{d\phi
}{dr}\right) ^{2}+V(\phi ,T)\right] , \label{s4}
\end{equation}
where an $O(4)$-symmetric configuration is assumed. The instanton
corresponds to the extremum of this action and is given by the equation
\begin{equation}
\frac{d^{2}\phi }{dr^{2}}+\frac{3}{r}\frac{d\phi }{dr}=\frac{\partial V}{
\partial \phi },  \label{eqprofile4}
\end{equation}
with the boundary conditions $d\phi /dr|_{r=0}=0$, $\lim_{r\rightarrow
\infty }\phi (r)=0$. At higher temperatures, the tunneling is affected by
thermal fluctuations. In the high-temperature limit we have $S=S_{3}(T)/T$
and  the prefactor in Eq.~(\ref{gamma}) is given by $A\simeq T^{4}\left[
S_{3}(T)/(2\pi T)\right] ^{3/2}$, where $S_{3}$ is the three-dimensional
instanton action \cite{linde}. Assuming spherical configurations, we have
\begin{equation}
S_{3}=4\pi \int_{0}^{\infty }r^{2}dr\left[ \frac{1}{2}\left( \frac{d\phi }{dr}
\right) ^{2}+V\right] ,  \label{s3}
\end{equation}
and the configuration of the nucleated bubble is given by the extremum of
$S_3$. Its equation is similar to Eq.~(\ref{eqprofile4}),
\begin{equation}
\frac{d^{2}\phi }{dr^{2}}+\frac{2}{r}\frac{d\phi }{dr}=\frac{\partial V}{
\partial \phi },  \label{eqprofile3}
\end{equation}
and the same boundary conditions apply. It is usual to consider the smaller
of $S_{4}$ and $S_{3}/T$  as an approximation for $S(T)$.

Analytic approximations for $S$ are sometimes considered (see, e.g.,
\cite{ah92}). These are useful to examine the parametric dependence as well
as the behaviour as the temperature varies. It can thus be seen that $S$
diverges at $T=T_{c}$, where the minima of the potential are degenerate, and
therefore the nucleation rate vanishes. Furthermore, $S$ vanishes at
$T=T_{0}$, where the potential barrier disappears, and hence near this limit
we have $\Gamma\sim T^4\sim v^4$. Due to the rapid variation of $S$ with $T$
and the exponential dependence of $\Gamma$, analytic approximations
generally introduce large errors in the nucleation rate. A better
approximation is to consider an expansion of $S$ around a certain
temperature as in Eqs.~(\ref{gammaexplin}-\ref{beta}), computing numerically
$S$ and its derivatives at that point. We shall compute $S$ by solving
Eqs.~(\ref{eqprofile4}) and (\ref{eqprofile3}) with the undershoot-overshoot
method and then integrating numerically the solution to obtain $S_{3}$ and
$S_{4}$.

The dimensionless quantity $S$ does not depend on the scale but only on the
form of the potential. For the model (\ref{potef}) this can be seen by
making the change of variables $R=vr,\Phi=\phi/v$ in
Eqs.~(\ref{s4}-\ref{eqprofile3}). With this rescaling, it is easy to see
that $S$ depends only on the dimensionless parameters $D,E,\lambda$ and the
ratios $A/v,T/v$. For concreteness, in this paper we will fix the values of
$m_{\phi }/v$, $E$ and $D$, and we will vary only the parameter $A/v$, which
plays a more determining role for the strength of the transition. We will
set $m_{\phi }=v/2$ (similar to the electroweak relation) and we will choose
a value of $E$ such that we already have $\phi _{c}/T_{c}=1$ for $A=0$. This
gives $\lambda =1/8$ and  $ E=1/16$. Finally, we set $D\simeq 0.44$, which
corresponds to a natural value for the latent heat\footnote{For $g_{\ast
}=100$ and $\phi_c/T_c=1$, this value of $D$ corresponds to $L=0.025\rho
_{R}$.}, $L\sim T_c^4$.

\begin{figure}[bt]
\centering
\epsfysize=7cm \leavevmode \epsfbox{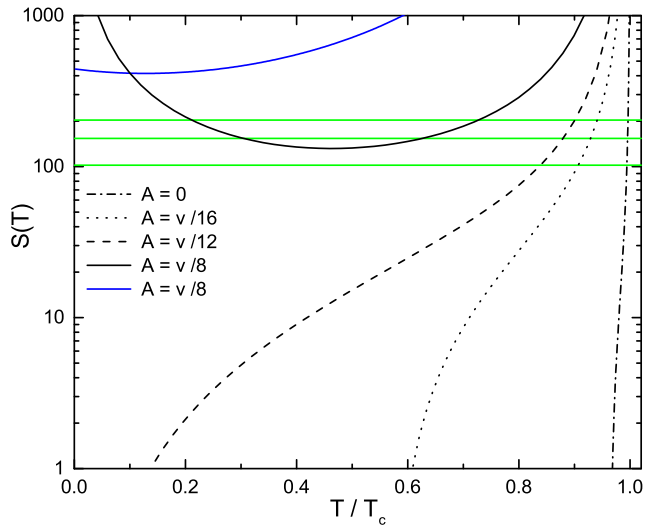}
\caption{The nucleation action as a
function of temperature, for different values of $A/v$. The blue curve corresponds
to $S=S_4$ and the rest to $S=S_3/T$.
The horizontal lines indicate the value $4\log (M_{P}/v)$ for $v=1\mathrm{MeV
}$ (upper line), $100\mathrm{GeV}$ (central line), and $10^{8}\mathrm{GeV}$
(lower line).}
\label{figs}
\end{figure}

In Fig.~\ref{figs} we plot $S$ as a function of $T/T_{c}$.
The rightmost curve corresponds to  $A=0$, and we see that for this kind of
phase transition (without tree-level cubic term) the temperature $T_{0}$ is
very close to the critical temperature. Thus, we have an extremely rapid
variation. The rest of the curves correspond to increasingly stronger phase
transitions. The spinodal temperature $T_{0}$ decreases as $A$ is increased,
and for $A=v/12$ (dashed line) we have $ T_{0}=0$. For higher values of
$A/v$ the potential barrier persists at zero temperature, and therefore the
action $S_{3}$ never vanishes. In this case, $S_{3}/T$ diverges at $T=0$
(black solid curve). This indicates that, for very low temperatures, bubble
nucleation by thermal activation becomes impossible. In that case, the
nucleation may occur by quantum tunneling, and we show also the action
$S_{4}$ (blue solid line).

Although  $S(T)$ does not depend on  $v$, the dynamics of nucleation does
depend on the scale through the expansion rate, which determines the rate at
which the universe cools down,
\begin{equation}
\frac{1}{T}\frac{dT}{dt}=-\sqrt{\frac{8\pi }{3}\frac{\rho _{+}(T)}{M_P^2}}.
\label{adexp}
\end{equation}
where $M_{P}=1.22\times 10^{19}\mathrm{GeV}$ is the Planck mass. As already
discussed, bubble nucleation will become appreciable when $\Gamma\sim
H^{4}$. For $T\sim v$, the value $\Gamma=H^4$ corresponds to $S=S_H\simeq
4\log (M_{P}/v)$. This value is generally large, provided that $v$ is a few
orders of magnitude below the Planck scale. Due to the logarithmic
dependence, for the wide range of scales $10^{8}\mathrm{GeV}
>v>1\mathrm{MeV}$ we have $100\lesssim S_H\lesssim
200$ (see the horizontal green lines in Fig.~\ref{figs}). For a given scale,
the value $S=S_H$ is reached at smaller temperatures for stronger phase
transitions. If $T_0^2\geq 0$ (dashed-dotted, dotted, and dashed curves),
then $S_{3}/T$ becomes arbitrarily small as $T$ decreases, and the condition
$\Gamma \sim H^{4}$ will eventually be reached. In contrast, when $S_{3}/T$
has a minimum (as in the black solid line), if the condition $ \Gamma \sim
H^{4}$ has not been fulfilled by the time this minimum is reached, then  it
will never be fulfilled ($\Gamma$ never gets high enough). Indeed, notice
that in such a case $S_{4}$ (blue solid line) is higher than $S_{3}/T$ until
the latter has become very high, and therefore the quantum tunnelling is
also very suppressed\footnote{This is a general feature and not a
particularity of our toy model (see, e.g., \cite{eknq08}).}.  The universe
will thus remain stuck in the false vacuum and enter an inflationary stage.
We are not interested in such a scenario\footnote{The quantum tunneling rate
may be important, for instance, if a hidden sector is in a metastable phase
at approximately zero temperature, and decays once the equality $H^4=\Gamma$
is reached (with $H$ depending on the temperature of the visible sector).
Such a scenario may lead to GWs via bubble collisions \cite{gkm16}.}, and
from now on we shall only consider cases in which the nucleation action is
given by $S(T)=S_{3}/T$.

\subsection{Bubble growth and global dynamics}

Once nucleated, a bubble begins to grow due to the pressure difference
between the two phases, $p_{-}(T)-p_{+}(T)=\mathcal{F}_{+}(T)-
\mathcal{F}_{-}(T)$. This initial driving force is affected by the local
reheating  (see, e.g., \cite{gkkm84,k85,kk86,eikr92,l94,ms09,ekns10,lm15}),
and thus depends on the values $T_{+}$ and $T_{-}$ of the temperature on
each side of the wall (the wall width is in general microscopic in
comparison with fluid profiles). Analytic expressions can be obtained by
using the bag EOS,
\begin{equation}
\mathcal{F}_{\pm }(T)=\varepsilon _{\pm }-a_{\pm }T^{4}/3, \label{bag}
\end{equation}
as an approximation for the free energy density in each phase. The model
(\ref{freeen}) has the form of the bag EOS for the $+$ phase, with
$\varepsilon _{+}=\rho _{V}$ and $a_{+}=\pi ^{2}g_{\ast }/90$. For the $ -$
phase, the approximation (\ref{bag}) can be used by defining
temperature-dependent parameters $ a_{-}=-(3/4)\mathcal{F}_{-}^{^{\prime
}}/T^{3}$, $\varepsilon_{-}= \mathcal{F}_{-}-T\mathcal{F}_{-}^{\prime }/4$
(see Ref.~\cite{ewur}). Using a linear approximation for the temperature
inside the wall, we obtain  the driving force \cite{urwalls}
\begin{equation}
F_{\mathrm{dr}}=\Delta \varepsilon (1-T_{+}^{2}T_{-}^{2}/T_{c}^{4}),
\label{fdr}
\end{equation}
where  $\Delta \varepsilon =\varepsilon _{+}-\varepsilon _{-}$. The
temperatures $T_{\pm}$ are related to the speeds of the incoming and
outgoing flows in the reference frame of the wall, which we denote $v_+$ and
$v_-$, respectively. We have (see, e.g., \cite{lm11})
\begin{equation}
\frac{a_{-}T_-^4}{a_{+}T_+^4}=\frac{v_{+}\gamma _{+}^{2}}{v_{-}\gamma _{-}^{2}}
\label{wme}
\end{equation}
and
\begin{equation}
v_{-}=\left( \frac{v_{+}\left( 1+\alpha_T \right) }{2}+\frac{\frac{1}{3}
-\alpha_T }{2v_{+}}\right) \pm \sqrt{\left( \frac{v_{+}\left( 1+\alpha_T \right)
}{2}+\frac{\frac{1}{3}-\alpha_T }{2v_{+}}\right) ^{2}-\frac{1}{3}},
\label{vmevma}
\end{equation}
where $\alpha_T =\Delta \varepsilon /(a_{+}T_{+}^{4})$.

For $0<v_{\pm}<1$, the relation (\ref{vmevma}) has four branches. In the
first place, the signs in front of the square root give two values of
$v_{-}$, either supersonic or subsonic, for a given $v_+$. Similarly, there
are two possible values of $v_+$ for a given $v_-$. Notice that the argument
of the square root is positive only for large enough or small enough $v_+$.
Since we are interested in very strong phase transitions, we need to
consider the branch which is compatible with fast walls, namely, that with
large $v_{+}$ and large $v_{-}$. These hydrodynamic solutions are called
weak detonations. In this case, the fluid outside the bubble is not affected
by the motion of the wall. Thus, the incoming flow velocity $v_{+}$ is given
by the wall velocity $v_{w}$, and the temperature $T_{+}$ is given by the
initial condition outside the bubble. These fluid equations are valid for a
steady state wall\footnote{We will not take into account the initial
acceleration stage, which is much shorter than the duration of the phase
transition provided that the temperature is a few orders of magnitude below
the Planck scale.} propagating at constant velocity, as well as for a
runaway wall which has reached the ultra-relativistic limit $ v_{w}\simeq
1$.

Besides the macroscopic effects of reheating, the wall motion is affected by
microphysics inside the wall
\cite{lmt92,t92,k92,a93,mp95,m00,js01,ms10,knr14,k15}. The departures from
equilibrium of the particle distributions result in a friction force. This
force is well understood in the non-relativistic (NR) and ultra-relativistic
(UR) limits. In the NR case, the friction grows linearly with the velocity,
while for $v_{w}\rightarrow 1$ it approaches a constant value \cite{bm09}.
We shall use the approximation \cite{ariel13}
\begin{equation}
F_{\mathrm{fr}}=\frac{\eta _{NR}\eta _{UR}\bar{v}}{\sqrt{\eta _{NR}^{2}\bar{v
}^{2}+\eta _{UR}^{2}(1-\bar{v}^{2})}},  \label{ffr}
\end{equation}
which interpolates between these regimes and depends on two free parameters
$\eta _{NR}$ and $\eta _{UR}$.  Here, $ \bar{v}$ is the average fluid
velocity in the wall frame, $\bar{v} =(v_{-}+v_{+})/2$. For small velocities
Eq.~(\ref{ffr}) gives $ F_{\mathrm{fr}}^{NR}\simeq \eta _{NR}\bar{v}$, while
in the UR limit we have $F_{\mathrm{fr} }^{UR}=\eta _{UR}$. The NR friction
coefficient $\eta _{NR}$ depends on the diffusion of particles which
interact with the wall. It is not directly related to the effective
potential, and we shall consider it as a free parameter.  On the other hand,
$\eta _{UR}$ can be obtained from $F_{ \mathrm{fr}}^{UR}=
F_{\mathrm{dr}}^{UR}- F_{\mathrm{net}}^{UR}$, where $F_{\mathrm{dr}}^{UR}$
is the UR limit of the driving force (\ref{fdr}) and $F_{\mathrm{net}}^{UR}$
is the UR limit of the net force. The latter is given by the mean field
value of the effective potential \cite{bm09}. For the model (\ref{potef}) we
have
\begin{equation}
F_{\mathrm{net}}^{UR}=-D(T^{2}-T_{0}^{2})\phi _{-}^{2}+A\phi _{-}^{3}-
\frac{\lambda }{4}\phi _{-}^{4}.  \label{fur}
\end{equation}
Notice that only for $A>0$ we can have $F_{\mathrm{net}}^{UR}>0$. If
$F_{\mathrm{net}}^{UR}$ is negative, it means that in the UR limit the
friction is higher than the driving force. In that case the UR regime cannot
be reached.

The steady state wall velocity is given by the equation
$F_{\mathrm{dr}}=F_{\mathrm{fr}}$, and can be computed using
Eqs.~(\ref{fdr}-\ref{fur}). Like for $S(T)$, the result does not depend on
the scale. The wall velocity is a function of the dimensionless ratios $\eta
_{NR}/T^4$ and $\eta _{UR}/T^4$, as well as of the parameter combination
$\alpha_T$. The latter is essentially given by the ratio of latent heat to
radiation energy density, $\alpha_T\sim L/\rho_R$, and contains the
information on hydrodynamics. For $F_{\mathrm{net}}^{UR}>0$, these equations
may give $v_{w}> 1$, indicating that the steady state condition cannot be
fulfilled and we have a runaway wall. In that case, we must set $v_{w}=1$,
and the value of $F_{\mathrm{dr}}-F_{\mathrm{fr}}$ will match the UR limit
(\ref{fur}).

This calculation of the wall velocity is valid for an isolated bubble. When
bubbles merge to form larger domains, the shapes of the walls, as well as
the fluid profiles, change. Although the above equations do not depend on
the wall shape\footnote{Surface tension effects in colliding bubbles can be
neglected, since the corresponding force is effective in very small scales
of time and length due to the large ratio $M_{P}/v$.}, the fluid profiles do
\cite{lm11}, and this may also change the boundary conditions. Nevertheless,
for detonations the temperature $T_+$ is not affected by the wall motion (in
contrast, for deflagrations the $+$ phase would be reheated). Inside the
bubbles we will have inhomogeneous reheating, but we do not expect this to
have a significant effect until the end of the phase transition. Thus, $T_+$
is only determined by the adiabatic expansion and we have $T_+\equiv T$,
with $T(t)$ given by Eq.~(\ref{adexp}). Notice that the scale factor is
given by a similar equation, since the square root in Eq.~(\ref{adexp}) is
just the expansion rate $H\equiv\dot a/a$. Actually, in the Friedmann
equation, the average energy density $ \bar{\rho}$ for a large volume
containing the two phases should be used. Nevertheless, the energy
difference between the two phases goes into reheating and fluid motions in
the $-$ phase, and the average energy density in that phase is approximately
given by $\rho _{+}(T)$.

At time $t$, the radius of a bubble which nucleated at time $t^{\prime }$ is
given by
\begin{equation}
R(t^{\prime },t)=\int_{t^{\prime }}^{t}dt^{\prime \prime }v_{w}(t^{\prime
\prime })\frac{a(t)}{a(t^{\prime \prime })},  \label{radio}
\end{equation}
where we have neglected the initial radius, which is a good approximation if
the phase transition takes place a few orders of magnitude below the Planck
scale. The fraction of volume occupied by the $+$ phase at time $t$ is given
by \cite{gt80,gw81}
\begin{equation}
f_{+}(t)=\exp \left[ -I(t)\right] ,  \label{fmas}
\end{equation}
with
\begin{equation}
I(t)=\int_{t_{c}}^{t}dt^{\prime }\,\Gamma (t^{\prime })\frac{
a(t^\prime)^{ 3}}{a(t)^{3}}\,\frac{4\pi }{3}R\left( t^{\prime },t\right) ^{3},
\label{I}
\end{equation}
where $t_{c}$ is the time at which $T=T_{c}$.
The exponential in Eq.~(\ref{fmas})
takes into account bubble overlapping, and the scale factors in
Eq.~(\ref{I}) take into account the dilution of nucleated bubbles. Since
bubbles only nucleate in the $+$ phase, the number of bubbles per unit time
per unit volume which are nucleated in a large volume containing regions of
both phases is given by the average nucleation rate
\begin{equation}
\bar{\Gamma}(t)=f_{+}(t)\Gamma (t).  \label{gammabar}
\end{equation}
Thus, as bubbles fill all space, their nucleation turns off. The number
density of bubbles at time $t$ is given by
\begin{equation}
n(t)=\int_{t_{c}}^{t}dt^{\prime }\bar{\Gamma}(t^{\prime })\frac{a(t^{\prime})^3
}{a(t)^{3}}.  \label{nb}
\end{equation}
We shall estimate the average distance between centers of nucleation by
\begin{equation}
d(t)=n(t)^{-1/3},
\end{equation}
while the mean radius is given by
\begin{equation}
\bar{R}(t)=\frac{1}{n(t)}\int_{t_{c}}^{t}dt^{\prime }\,\bar{\Gamma}(t^{\prime })\frac{
a(t^{\prime})^3}{a(t)^{3}}\,R(t^{\prime },t).  \label{rm}
\end{equation}
The distribution of bubble sizes is given by
\begin{equation}
\frac{dn}{dR}(t)=\frac{f_{+}(t_{R})\Gamma (t_{R})}{v_{w}(t_{R})}\frac{
a(t_{R})^{4}}{a(t)^{4}},  \label{dndR}
\end{equation}
where $t_{R}$ is the time at which the bubble of radius $R$ was nucleated,
which is obtained by inverting Eq.~(\ref{radio}) for $t^{\prime }$ as a
function of $R$ and $t$.

The temperature at which the phase transition ``effectively'' begins should
be close to the temperature $T_{H}$ at which the equality $\Gamma =H^{4}$ is
fulfilled, which, as already mentioned, is roughly given by the value
$S_H\simeq 4\log (M_{P}/v)$. More precisely, $T_{H}$ and $S_H$ are
determined by the equation
\begin{equation}
S(T_{H})-(3/2)\log [S(T_{H})/(2\pi )]=4\log \left[ T_{H}/H(T_{H})\right] .
\label{tH}
\end{equation}
(with $S=S_3/T$). However, a sensible definition of the beginning of the
transition should involve the dynamics of nucleation. It is usual to define
the ``onset of nucleation'' by demanding that there is already one bubble in
a Hubble volume. We thus define the corresponding time $t_{N}$ and
temperature $T_N$ by the condition $nH^{-3}=1$. 
Using instead the fraction of volume in the $+$ phase as a measure of the
progress of the phase transition, we may define the initial time as that
$t_{I}$ at which the variation of this quantity becomes noticeable, e.g., by
the condition $f_{+}(t_{I})=0.99$. Below we shall compare these definitions.
With the quantity $f_+(t)$ we may define a few more reference points in the
evolution of the phase transition. A representative moment at which there
are already many bubbles in contact is when the percolation threshold is
reached. This is the time at which there is an infinite cluster of connected
bubbles. For spherical bubbles of equal size $ R $ and number density $n$,
this happens when $1-\exp (-I)\simeq 0.29$, with $I=(4\pi /3)R^{3}n$
\cite{perco}. Therefore, we define the percolation time $t_{P}$ by
$f_{+}(t_{P})=0.71$. Another characteristic time which is often considered
is the time $t_{E}$ at which the fraction of volume in the false-vacuum
phase has fallen to $f_{+}(t_{E})=1/e$. Equivalently, we have $I(t_{E})=1$.
Notice that the value $f_{+}(t)=0$, corresponding to the completion of the
phase transition, is approached asymptotically for $t\rightarrow \infty $ in
the approximation (\ref{fmas}-\ref{I}). However,  in general $f_+$ becomes
very small in a time of the order of $t_E-t_I$. Physically, this indicates
that the phase transition actually completes. We shall define a
representative time $ t_{F} $ at which $f_+$ has become small by
$f_{+}(t_{F})=0.01$. We shall denote the values of the various quantities at
all these reference points with corresponding subscripts $H$, $I$, $N$, $P$,
$E$ and $F$.

The dynamics of the phase transition will depend mainly on the parameters of
the effective potential\footnote{As discussed above, we will consider
$E=0.0625$, $D\simeq 0.44$, $m_{\phi }/v=0.5$, and vary $A/v$.} $V(\phi,T)$.
The Hubble rate and the wall velocity depend also on the radiation energy
density, and hence on the number of degrees of freedom of the plasma.
Besides, $H$ depends on the scale. For concreteness, we shall set $g_*=100$
and  $v= 250 \mathrm{GeV}$ (i.e., electroweak scale values; as discussed
above, for other scales there will not be qualitative differences). For the
wall velocity we need to consider a specific value for the non-relativistic
friction parameter, and we shall set $\eta _{NR}=0.25T_{c}^{4}$. The precise
value of $\eta_{NR}$ is not very important for the strong phase transitions
we are interested in, since the friction will be dominated by the UR
coefficient, which is determined by Eq.~(\ref{fur}).

\begin{figure}[tb]
\centering
\epsfysize=5.5cm \leavevmode \epsfbox{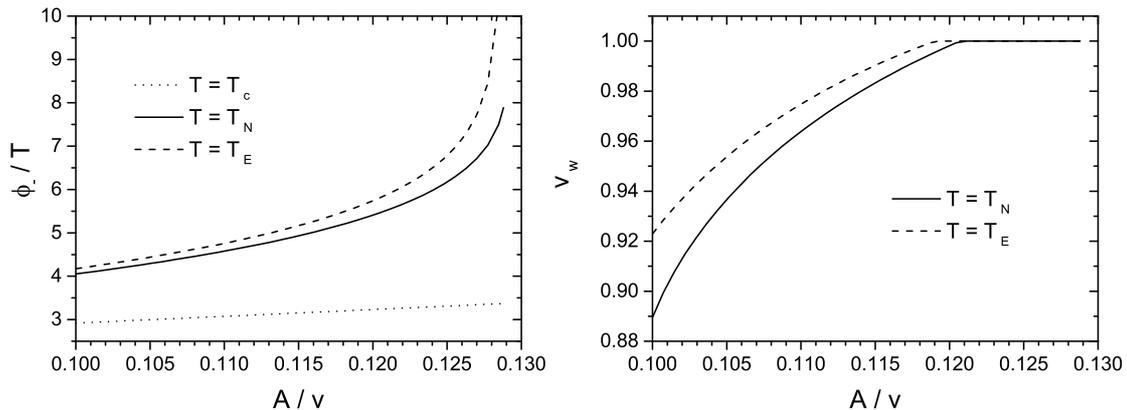}
\caption{The wall velocity and order parameter at different temperatures.}
\label{figvw}
\end{figure}

In the left panel of Fig.~\ref{figvw} we show the order parameter at some
reference temperatures, for a range of $A/v$ in which the phase transition
is very strong.
We have  $\phi_c/T_c\gtrsim 3$, and we see that by the onset of nucleation
the order parameter is significantly larger. This is due to a combination of
two effects. Firstly, the minimum $\phi_-$ is closer to the zero-temperature
value (which is larger than $\phi_c$), and secondly the temperature has
decreased from $T_c$. We also see that the strength of the transition
increases very rapidly with $A/v$, so for $A/v\simeq 0.13$ the transition
becomes so strong that $T_N$ vanishes. In the right panel of
Fig.~\ref{figvw} we show the wall velocity. In this parameter range we have
detonations and runaway walls, while for lower values of $A/v$ we have
deflagrations. Thus, there are detonations for only about 15\% of the total
range of $A/v$ and runaway walls for only 8\% of the range. These
proportions are typical (see, e.g.~\cite{ewur}), although they depend of
course on the rest of the parameters. Notice that for this parameter range
the wall velocity does not change significantly during the development of
the transition.

\begin{figure}[bt]
\centering
\epsfysize=7cm \leavevmode \epsfbox{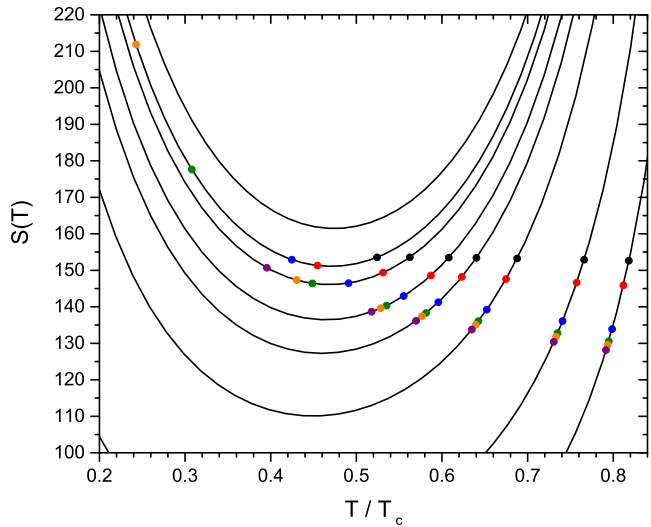}
\caption{The curves of $S(T)$ for (from bottom to top) $A/v=0.1$, $0.11$, $0.12$,
$0.124$, $0.126$, $0.128$, $0.129$, and $0.131$. The dots correspond to the reference points
$H$ (black), $N$ (red), $I$ (blue), $P$ (green), $E$
(orange), and $F$ (purple),  for $v=250
\mathrm{GeV}$.}
\label{figpoints}
\end{figure}

In Fig.~\ref{figpoints} we show the reference points on the curves of
$S(T)$, for several values of $A/v$.
In the upper curve ($A/v=0.131$) the equality $\Gamma =H^{4}$ is never
reached and the system remains stuck in the false vacuum. In the rest of the
curves this equality is reached for $S\simeq 153$ (approximately the same
for all the curves), which agrees with the approximation $ S_{H}\simeq 4\log
(M_{P}/v)$. This point is indicated with a black dot on each curve. The
equality $n=H^{3}$ is reached for values of $S$ around $146$-$151$ (red
dots). The little variation of $S_{N}$  indicates that the initial dynamics
of nucleation does not depend much on the model. Notice that the
supercooling is quite strong, $T_{N}/T_{c}< 0.8$. In comparison, for $A=0$
(for which $\phi _{c}/T_{c}=1$) we have $T_{N}/T_{c}\simeq 0.99$. The values
of $S$ at the points $I,P,E,F$ (which are characterized by specific values
of the fraction of volume) are more dependent on the global dynamics of the
phase transition and, hence, on the model. Notice also that in the second
curve from the top ($A/v=0.129$), very small temperatures are reached during
the phase transition. This happens because the minimum of $S$ is surpassed
very soon after $t=t_{H}$, and the nucleation rate starts to decrease even
before there is one bubble per Hubble volume. In spite of this, we see that
the various reference points are reached (the point $F$ lies out of the
range of the figure, $T_F\simeq 0.1$).

\begin{figure}[tbp]
\centering
\epsfxsize=15cm \leavevmode \epsfbox{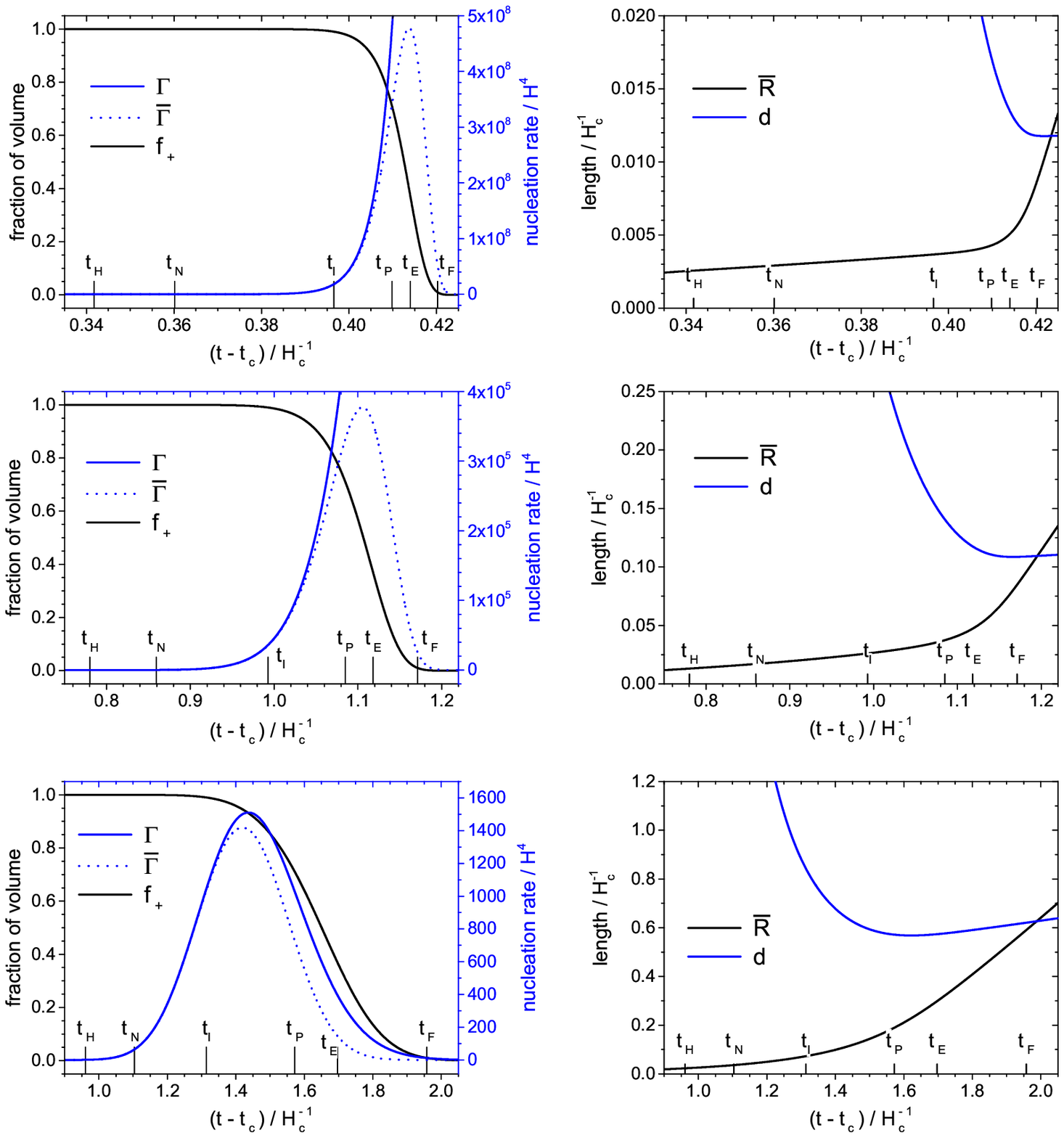}
\caption{The evolution of several
quantities, for the cases $A/v=0.11$ (top), $A/v=0.126$
(middle) and $A/v=0.128$ (bottom). The left panels show the fraction
of volume $f_{+}$, the nucleation rate $\Gamma $ and the average nucleation
rate $\bar{\Gamma}$. The right panels show the average radius $\bar R$ and
the bubble separation $d$.}
\label{figevol}
\end{figure}

In Fig.~\ref{figevol} we show the evolution of some quantities for a few of
the cases considered in Fig.~\ref{figpoints}. We also indicate the reference
times. The plots on the top correspond to $A/v=0.11$. The significant amount
of supercooling can be appreciated in the fact that the time from $t_{c}$ to
$t_{N}$ is quite longer than the time from $t_{N}$ to $t_{F}$, in which the
phase transition actually develops. We also see (left panel) that the
interval $ [t_{I},t_{F}]$ in which the fraction of volume varies in also
quite shorter than the latter. Furthermore, most bubbles nucleate in this
short interval near the end of the transition.
In the top right panel of Fig.~\ref{figevol}, we see that the average
distance between bubble centers and the average bubble size become
comparable by the end of the transition, as expected. The bubble separation
has initially a very rapid variation, since $\Gamma $ is increasingly high,
but becomes constant at $ t=t_{F}$, as bubbles cease to nucleate. We also
see that the average radius grows very slowly during most of the phase
transition, even though all bubbles grow with a velocity $v_{w}\simeq 1$.
The small average is caused by the constant nucleation of vanishingly small
bubbles. This effect disappears for $t\simeq t_{F}$, as the space where
bubbles nucleate is considerably reduced, and we have $d\bar{R}/dt\simeq
v_{w}$.

Although we will be mostly interested in higher values of $A/v$, it is worth
mentioning that the evolution will be qualitatively similar for weaker phase
transitions, provided that bubbles expand as detonations. Quantitatively,
the variation of $f_{+}$ and $\bar{\Gamma}$ will be even more concentrated
at the end of the phase transition, since the function $S(T)$ will have a
steeper slope (see Figs.~\ref{figs} and \ref{figpoints}). Thus, if $T_N$ is
close to the critical temperature, the nucleation rate will grow at a
dramatically increasing rate. In such a case, the dynamics is dominated by
the nucleation of bubbles, rather than by their growth. In contrast, for
stronger phase transitions we have more supercooling, the temperature gets
closer to the minimum of $S$, and the nucleation rate varies more slowly.

\begin{figure}[bt]
\centering
\epsfysize=5.2cm \leavevmode \epsfbox{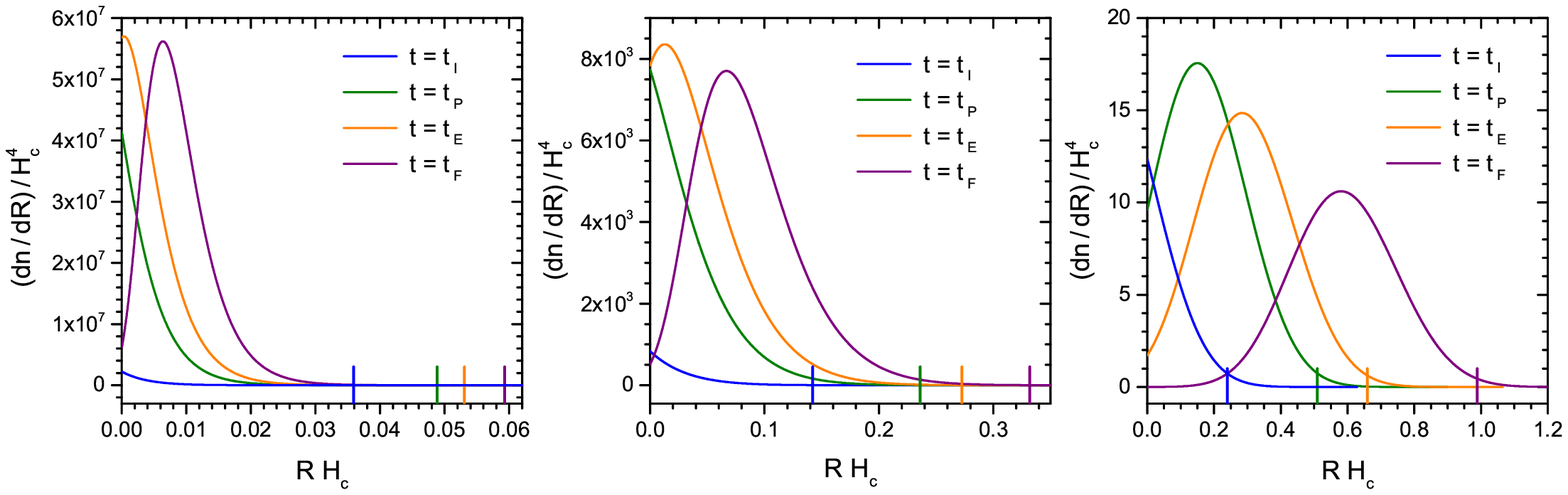}
\caption{The bubble size distribution $
dn/dR$ at several times, for $A/v=0.11$ (left), $A/v=0.126$
(center), and $A/v=0.128$
(right). The long ticks indicate the radius of the bubbles nucleated at $
t=t_{N}$.}
\label{figdistrib}
\end{figure}

In the left panel of Fig.~\ref{figdistrib} we plot the distribution of
bubble sizes for the same case $A/v=0.11$, at the times $t_{I}$, $t_{P}$,
$t_{E}$ and $t_{F}$. We see that, for $t\leq t_{E}$, vanishingly small
bubbles are more abundant. This is due to the rapid growth of the nucleation
rate. On the other hand, for $t>t_{E}$ the maximum of the distribution
separates from $R=0$, indicating that the average nucleation rate has
decreased. Notice that the width of this distribution is essentially the
same as that of $\bar{\Gamma}$, as expected from Eqs.~(\ref{gammabar}) and
(\ref{dndR}). We also indicate in Fig.~\ref{figdistrib} the radius of the
\textquotedblleft largest\textquotedblright\ bubbles, $R(t_{N},t_{i})$, at
each reference time $t_{i}$. We see that, for this case, the average size is
generally quite smaller than the largest bubble.

The second row of plots in Fig.~\ref{figevol} corresponds to the case
$A/v=0.126$. In this case  we have an important amount of supercooling ($S$
is quite close to the minimum, as can be seen in Fig.~\ref{figpoints}).
Therefore, we have a larger  time $t_{H}-t_c$. Besides, due to the slower
growth of $\Gamma $ the development of the phase transition is slower, and
at the end of it we have fewer and larger bubbles. On the other hand, the
variation of $f_{+}$ begins \emph{relatively} sooner (i.e., closer to
$t_{N}$ in the interval $[t_{N},t_{F}]$). As a consequence, the curve of
$\bar{\Gamma}$ is less concentrated. This is reflected in the bubble size
distribution  (see the central panel of Fig.~\ref{figdistrib}), which is
wider in relation to the largest bubble size $ R(t_{N},t)$. Due to the
longer phase transition, the effect of the scale factor begins to be
noticeable in the curve of $d$ as well as in the evolution of $dn/dR$.

The third row of Fig.~\ref{figevol} and the right panel of
Fig.~\ref{figdistrib} correspond to the case $A/v=0.128$. In this case the
minimum of $S$ is crossed during the phase transition, and the nucleation
rate turns off even in the $+$ phase. Thus, the effective rate
$\bar{\Gamma}$ drops to zero more quickly than $f_{+}$. Besides, since
$\Gamma $ reaches a maximum, it varies much more slowly than in the previous
cases. This can be seen e.g. in the curves of $\bar{\Gamma}$ and the bubble
size distribution, which are quite wider. The dynamics is no longer
dominated by bubble nucleation, but also depends on the bubble growth and on
the expansion rate of the universe. The effect of bubble growth can be seen
in the curve of $\bar{R}$, which reaches the behavior $\propto v_{w}t$
sooner than in the previous cases (i.e., closer to $t_{N}$). This can be
appreciated also in the evolution of the bubble size distribution, which at
$t=t_{P}$ has already a non-vanishing maximum. The effect of the expansion
of the universe is reflected, e.g., in the distance $d$ between bubble
centers. As soon as bubble nucleation slows down, $d$ stops decreasing and
begins to grow. This shows that the variation of the scale factor during the
phase transition is not negligible. The effect is more evident in the
distribution of bubble sizes, where the height of the curve of $dn/dR$
decreases with $t$ due to the dilution of the number density as $n\propto
a^{-3}$.

Although these three examples correspond to very strong phase transitions,
we see that the dynamics  becomes qualitatively different near the minimum
of $S$. The difference is particularly evident if we compare  the leftmost
and rightmost panels in Fig.~\ref{figdistrib}. The bubble size distribution
is not only quantitatively and qualitatively different, but it also evolves
quite differently.

\begin{figure}[bt]
\centering
\epsfysize=5.5cm \leavevmode \epsfbox{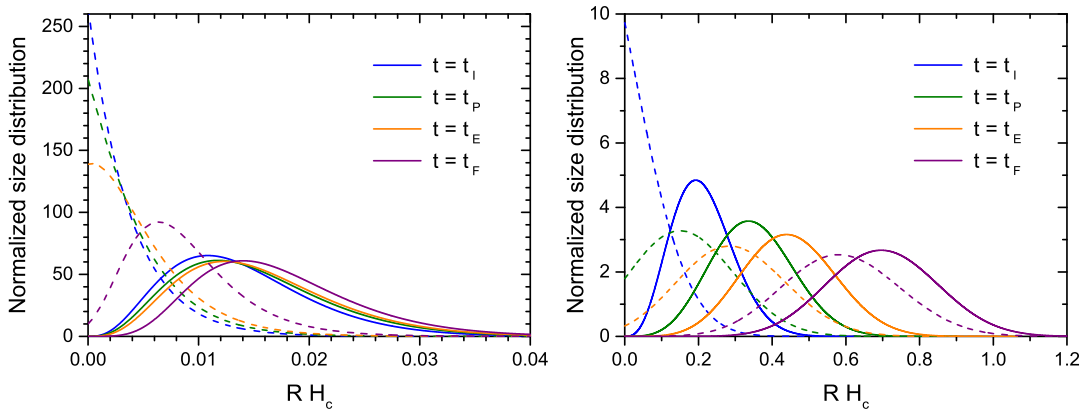}
\caption{The bubble size distributions $(R^3dn/dR)/(\int R^3dn)$ (solid lines)
and $(dn/dR)/n$ (dashed lines)
at different times as in Fig.~\ref{figdistrib}, for the cases
$A/v=0.11$ (left) and $A/v=0.128$ (right).}
\label{figndistrib}
\end{figure}

In discussing GW generation it is more appropriate to look at the
volume-weighted bubble distribution, $R^3 dn/dR$, since the energy injected
into the walls of a bubble and into the surrounding fluid is proportional to
the volume of the bubble. In Fig.~\ref{figndistrib} we show the
volume-weighted distribution together with the radius distribution, for the
cases $A/v=0.11$ and $A/v=0.128$.
For a better comparison, we have divided the distributions by their
integrals. The volume-weighted distribution vanishes at $R=0$, and therefore
is peaked at a larger radius.  For this distribution we still observe a
different behavior in the two examples. Indeed, in the left panel the
distribution is almost unchanged during the phase transition, while in the
right panel it moves fast to larger $R$ and it also spreads.

\section{Analytic approximations} \label{anap}

Analytic approximations are useful for a better understanding of the physics
as well as for specific applications. In particular, analytic approximations
for the nucleation rate are often used in simulations of the phase
transition. If the duration  is short enough, then the linearization of
$S(t)$ is a good approximation and we may use a nucleation rate of the form
(\ref{gammaexplin}) \cite{h83,tww92}. In this approximation, the constant
$\beta ^{-1}$ will set the time scale for the dynamics. One expects that the
approximation will be good if this time scale is shorter than that for the
temperature variation, which is given by $H^{-1}$. Therefore, the
approximation will be valid for $\beta /H\gg 1$. More concretely,  consider
the second-order expansion of $S(t)$ about a certain $t_{\ast }$,
\begin{equation}
S(t)\simeq S(t_*)-\beta(t-t_*)+\left(\alpha^2+\frac{\beta (H-\dot H/H)}{2}\right)(t-t_*)^2,
\label{quadexp}
\end{equation}
with $\dot H=dH/dt$,
\begin{equation}
\beta= H\,T\frac{dS}{dT},\qquad  \alpha^2=H^2\,\frac{T^2}{2}\frac{d^2S}{dT^2},
\end{equation}
and all the quantities are evaluated at $t=t_*$. The linear approximation
will be valid if we can neglect the last term in (\ref{quadexp}), which
implies the conditions
\begin{equation}
 \dot H(t-t_*)\ll 2H,\qquad H(t-t_*)\ll 2, \qquad \alpha^2(t-t_*)\ll \beta. \label{linval}
\end{equation}
The first condition requires a small variation of $H$ in the time $(t-t_*)$,
while the second condition is just the requirement of a short time compared
to $H^{-1}$. Since in general $\dot H/H\sim H$, these are equivalent.
Assuming that they are fulfilled, the third condition requires that the
coefficient $\beta=-dS/dt$ does not change too much if it is evaluated at
the time $t$ instead of $t_*$. Indeed, in a time $\Delta t$ we have  a
variation $\Delta\beta\simeq 2\alpha^2\Delta t$. Hence, the last condition
in (\ref{linval}) just demands that $\Delta \beta\ll\beta$.

For the exponential rate to be a good approximation, Eqs.~(\ref{linval})
should be valid at least for the characteristic time $t-t_*\sim\beta^{-1}$.
Therefore, we have
\begin{equation}
\beta\gg H,\qquad \beta\gg\alpha  . \label{condba}
\end{equation}
 The second condition does not depend on $H$ and imposes a relation
between the temperature derivatives of $S(T)$. As we have seen, the phase
transition in general takes place for high values of $S$, namely, $S\sim
4\log (M_{P}/v)\gg 1$. Hence, for natural values of the derivatives we have
$TdS/dT\sim T^{2}d^{2}S/dT^{2}\sim S$, which give $\beta /H\sim S$, $\alpha
/H\sim \sqrt{S}$. Both quantities are generally large, and so is the ratio
$\beta/\alpha\sim\sqrt{S}$. Therefore, the two conditions (\ref{condba})
will be fulfilled in most cases. Of course, there is a special case in which
this argument breaks down, namely, when the temperature gets too close to
the value $T_m$ corresponding to the minimum of $ S(T)$, at which $\beta $
vanishes. For the case considered in Figs.~\ref{figpoints} to
\ref{figndistrib} ($v=250\mathrm{GeV}$) we have $S\sim 150$. Table
\ref{tabbeta} shows the values of $\beta/H$ at several times $t_*$ for the
specific cases considered in Figs.~\ref{figevol} to \ref{figndistrib}. The
value of $\alpha/H$ at the minimum is also shown. In all these cases we have
$\alpha/H\sim 10$, while $\beta/H$ is roughly $\sim 10^2$. However, the
latter will be higher for weaker phase transitions, and it becomes small and
negative for $t_*$ very close to the minimum (cf. Fig.~\ref{figpoints}).
\begin{table}[bth]
\centering
\begin{tabular}{|c|c|c|c|c|c|c|c|} \hline $A/v$ & $\beta/H|_{T_H}$ & $\beta/H|_{T_N}$
& $\beta/H|_{T_I}$ & $\beta/H|_{T_P}$ & $\beta/H|_{T_E}$ & $\beta/H|_{T_F}$ & $\alpha/H|_{T_m}$ \\
\hline 0.11 & 583 & 525 & 435 & 408 & 400 & 389 & 8.7 \\
\hline 0.126 & 157 & 123 & 82 & 60 & 53 & 43 & 12.9 \\
\hline 0.128 & 93 & 56 & 18 & -15 & -28 & -51.3 & 13.5 \\ \hline
\end{tabular}
\caption{The values of $\beta/H$ at the reference points, and the value
of $\alpha/H$ at the minimum of $S$, for the cases considered in Fig.~\ref{figevol}.}
\label{tabbeta}
\end{table}

As $T$ approaches $T_m$, the second condition in Eq.~(\ref{condba}) will be
violated before the first one (since in general we have $\alpha\gg H$).
Thus, the linear approximation will break down when the value of
$\beta(T_*)$ becomes comparable to that of $\alpha(T_*)$. This establishes a
rough criterion for the validity of this approximation, namely,
$\beta(T_*)\gtrsim$ $\alpha(T_*)$, and this condition will only break down
near the minimum. Thus, we expect the exponential rate to be a good
approximation for $\beta(T_*)\gtrsim \mathrm{a \ few} \, \alpha(T_m)$.

If $T$ gets too close to $T_m$,  we may set $t_*=t_m$ in
Eq.~(\ref{quadexp}), and we obtain a gaussian approximation for the
nucleation rate,
\begin{equation}
\Gamma (t)=\Gamma _{m}e^{-\alpha ^{2}(t-t_{m})^{2}},  \label{gammagauss}
\end{equation}
with $\Gamma _{m}=\exp [-S(T_{m})]$. This quadratic approximation for $S(t)$
will also break down for $t$ far enough from $t_m$. In particular, we expect
that it will break down for $|t-t_m|\gtrsim H^{-1}$. Therefore, if the
temperature $T_m$ is not reached during the phase transition, in order to
use the approximation (\ref{gammagauss}) we must have at least $S(T)$ close
enough to $S(T_m)$. On the other hand, if temperatures below $T_m$ are
reached, then Eq.~(\ref{gammagauss}) can be safely used, even if the phase
transition takes a long time to complete, since $\Gamma$ will be negligible
after a time $\sim \alpha^{-1}$,  and, as we have seen, $\alpha ^{-1}\sim
(\sqrt{S}H)^{-1}\ll H^{-1}$.

\subsection{Exponential nucleation rate}

As can be appreciated in Fig.~\ref{figpoints}, in many cases the function
$S(T)$ is approximately linear during the phase transition or at least in
the relevant interval $[t_I,t_F]$ (where $f_{+}$ varies and the size
distribution is formed). This can also be seen in the first row of
table~\ref{tabbeta}, where $\beta$ varies by a 10\% between $t_I$ and $t_F$.
In such cases, the exponential nucleation rate (\ref{gammaexplin}) should be
a good approximation. Besides, as we have seen, we may also assume a
constant wall velocity. With these assumptions
Eqs.~(\ref{radio})-(\ref{dndR}) can be solved analytically \cite{tww92}. We
shall also neglect the evolution of the scale factor for simplicity,
although this is not necessary to obtain analytic results. Besides, when
comparing a time interval with the characteristic cosmic time $H^{-1}$, we
shall assume a constant $H$ in that interval, which is consistent with a
constant $\beta$. We shall later compare these approximations with the
numerical results obtained above.

With these approximations the radius is just given by $R(t^{\prime
},t)=v_{w}(t-t^{\prime })$, and taking, with little error, $t_{c}\rightarrow
-\infty $ in Eq.~(\ref{I}), we obtain \cite{tww92}
\begin{equation}
I(t)=I_{\ast }e^{\beta (t-t_{\ast })},  \label{Iexp}
\end{equation}
where $I_{\ast }=8\pi v_{w}^{3}\Gamma _{\ast }/\beta ^{4}$. Thus, we have
$I(t)=(8\pi v_{w}^{3}/\beta ^{4}) \Gamma(t)$. With $ f_{+}=e^{-I}$, this
gives
\begin{equation}
\bar{\Gamma}(t)=\Gamma _{\ast }e^{\beta (t-t_{\ast })}\exp [-I_{\ast
}e^{\beta (t-t_{\ast })}],  \label{avgaexp}
\end{equation}
which is easily integrated to obtain the number density,
\begin{equation}
n(t)=\frac{\beta ^{3}}{8\pi v_{w}^{3}}\left[ 1-f_{+}(t)\right] .
\label{nexp}
\end{equation}
Actually, at the beginning of the transition we may ignore the factor
$e^{-I}$ in Eq.~(\ref{avgaexp}), and we obtain (for $f_{+}\simeq 1$)
\begin{equation}
n=\Gamma (t)/\beta .  \label{niniexp}
\end{equation}
The distribution of bubble sizes is given by
\begin{equation}
\frac{dn}{dR}(t)=v_{w}^{-1}\,{\bar{\Gamma}(t-R/v_{w})},  \label{dndRexp}
\end{equation}
and for the average radius we have the expression
\begin{equation} \label{rmap}
\bar{R} =\frac{v_{w}}{n}\int_{t_{c}}^{t}dt^{\prime }\bar{\Gamma}(t^{\prime
})(t-t^{\prime }) =v_{w}\int_{t_{c}}^{t}dt^{\prime }\frac{n(t^{\prime
})}{n(t)}.
\end{equation}
In the last equality  we have used the relation $\bar{\Gamma}
=dn/dt^{\prime}$.

\begin{figure}[bt]
\centering
\epsfysize=5cm \leavevmode \epsfbox{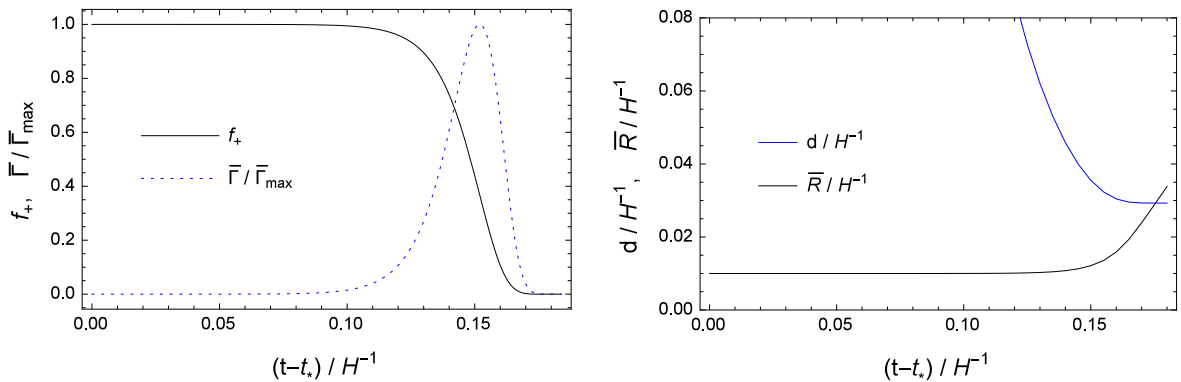}
\caption{The evolution of $f_{+}$, $\bar{
\Gamma}$, $d$ and $\bar{R}$ for the approximations
(\protect\ref{Iexp})-(\protect\ref{rmap}),
for $\Gamma _{\ast }=H^{4}$ ($t_{\ast }=t_{H}$),
$\protect\beta /H=100$, and $v_{w}=1$.}
\label{figaplin}
\end{figure}

To see the qualitative behavior of this approximation, it is convenient to
set  $v_{w}=1$ and choose the value $\Gamma _{\ast }=H^{4}$ (corresponding
to  $t_{\ast }=t_{H}$), so that, if time and distance are in units of
$H^{-1}$, all the above expressions depend on the single parameter
$\beta/H$. In Fig.~\ref{figaplin} we plot some of these quantities  for
$\beta /H=100$.
As expected, these plots show the qualitative behavior of the first row of
Fig.~\ref{figevol}, even though the present value of $\beta (t_H)$ is closer
to that corresponding to the third row (see table~\ref{tabbeta}). There is a
qualitative difference, though, between Fig.~\ref{figevol} and
Fig.~\ref{figaplin}, namely, that for most of the phase transition the
average radius remains constant in the latter, while in the former there is
a small slope. This difference is mainly due to the variation of $\beta $,
which is neglected in the present approximation.

More quantitatively, at the beginning of the phase transition we have
$\bar\Gamma\simeq\Gamma$ and the first integral in Eq.~(\ref{rmap}) gives
$\bar{R}(t)=v_{w}/\beta $. Moreover, this result does not change to first
order in $ I$. Indeed, from Eq.~(\ref{nexp}), the second integral in
(\ref{rmap}) gives $\bar R\simeq v_{w}\int_{t_{c}}^{t}dt^{\prime
}I(t^{\prime })/I(t)+\mathcal{O}(I^2)$. This gives $\bar R(t)=v_{w}/\beta+
\mathcal{O}(1-f_{+})^2$, which will remain constant until $f_{+}$ departs
appreciably from $f_{+}=1$. At first sight, such a constant average seems in
contradiction with the fact that, according to Eq.~(\ref{dndRexp}), the
bubble size distribution moves with velocity $v_{w}$ without changing its
shape. However, part of this curve falls into negative sizes. It is easy to
see that the peak is at $ R_{p}(t)=v_{w}(t-t_E)$. Therefore, we have $
R_{p}(t)<0$ for $t<t_E$. This means that before $t=t_{E}$ the  physical
distribution actually decreases from $R=0$ (in agreement with the left panel
of Fig.~\ref{figdistrib}). As a consequence, $\bar{R}$ will remain small
until $t= t_E$ (and by this time the phase transition is fairly advanced).
On the other hand, for $t\gtrsim t_{F}$ we have $\bar\Gamma\simeq 0$ and the
first integral in (\ref{rmap}) gives $ \bar{R}\simeq
\bar{R}(t_{F})+v_{w}(t-t_{F})$, as observed in the figures.

In contrast, since the volume-weighted  distribution vanishes at $R=0$, its
average radius is close to its maximum $R_p'$. The latter is given by
$R'_p=x(t)\,v_w/\beta$, where $x$ is the solution of the equation
$(x-3)e^x=xI(t)$. This peak is initially at $R'_p=3\,v_w/\beta$ (for $I=0$)
and moves to $R'_p\simeq 3.14\, v_w/\beta$ at $t=t_E$, while at $t=t_F$ we
have $R_p'\simeq 3.49\, v_w/\beta$. This small variation is in agreement
with the left panel of Fig.~\ref{figndistrib}. Another important distance
scale is the bubble separation $d=n^{-1/3}$. According to
Eq.~(\ref{niniexp}), $d$ initially reflects the rapid variation of $\Gamma$.
However, near the end of the transition $n$ reaches a constant value
$n={\beta ^{3}}/({8\pi v_{w}^{3}})$, and we have $d=(8\pi)^{1/3}v_w/\beta$.

For the exponential rate it is straightforward to obtain the relations among
the reference times defined in the previous section. At $t=t_H$ we have
$f_{+}=\exp (-8\pi v_{w}^{3}H^{4}/\beta ^{4})$, which in the usual case
$H/\beta \ll 1$ may be written as $f_{+}(t_{H})\simeq 1-8\pi
v_{w}^{3}(H/\beta)^{4}$ (in most cases the second term is actually
negligible). Then, we have $ n(t_{H})\simeq H^{4}/\beta $, which indicates
that the number of bubbles per Hubble volume is very small,
$n(t_{H})H^{-3}\simeq H/\beta \ll 1$. Therefore, the time $t_{H}$ is always
earlier than the onset of nucleation. At the time $t_N$ we may use
Eq.~(\ref{niniexp}), which gives
\begin{equation}
\Gamma (t_{N})/H^{4}= \beta /H.  \label{gatnlin}
\end{equation}
Comparing with $\Gamma (t_{H})/H^{4}=1$, we obtain
\begin{equation}
t_{N}-t_{H}\simeq \beta ^{-1}\log (\beta /H).  \label{tnexp}
\end{equation}
Notice that at the time $t_{N}$ we  have $ I(t_N)=8\pi v_{w}^{3} (H/\beta)^3
\ll 1$, i.e., $f_{+}\simeq 1$. On the other hand, the subsequent times
$t_i$, $ i=I,P,E,F$, correspond to specific (larger) values $I(t_i)=I_i$,
with $I_{i}=0.01$, $0.34$, $1$, and $4.6$. We thus have
$\Gamma(t_i)=I_i\beta^4/(8\pi v_{w}^{3})$. Comparing with (\ref{gatnlin}),
we have, for example,
\begin{equation}
t_{I}-t_{N}\simeq \beta ^{-1}\log \frac{(\beta /H)^{3}I_{I}}{8\pi v_{w}^{3}}.
\label{titnexp}
\end{equation}
For the last four points the time differences are given by $
t_{i}-t_{j}=\beta ^{-1}\log (I_{i}/I_{j})$. For instance, for the interval
in which $f_{+}$ varies we have
\begin{equation}
t_{F}-t_{I}\simeq 6\beta ^{-1}.  \label{tftiexp}
\end{equation}
If we take the width of $\bar\Gamma(t)$ as $t_F-t_I$, then the width of the
distribution of bubble sizes is $\Delta R\simeq  6v_{w}\beta ^{-1}$. We see
that all the time intervals are of order $\beta ^{-1}$, but for the time
from $ t_{H}$ to $t_{I}$ we have $\log (\beta /H)$ enhancements. In
contrast, the time in which the fraction of volume varies is fixed in units
of $\beta ^{-1}$. This difference becomes important for $\beta /H\gg 1$,
i.e., for weaker phase transitions. In the interval $[t_I,t_F]$, the time
$t_{E}$ is particularly interesting, since it corresponds to the moment of
maximum bubble nucleation, as can be easily seen by differentiating
Eq.~(\ref{avgaexp}). This agrees with the first case considered in
Fig.~\ref{figevol}, and is also in good agreement with the second case.

\subsection{Gaussian nucleation rate}

The gaussian approximation (\ref{gammagauss}) will be more appropriate to
describe a phase transition like the one shown in the last row of
Fig.~\ref{figevol}. Neglecting the variation of the wall velocity, the
evolution of the phase transition can be solved again analytically, although
the expressions are more involved. For the moment we will also ignore the
variation of the scale factor.

At the beginning of the phase transition the number density of bubbles is
just given by the integral of $ \Gamma $, and we have (for $f_{+}\simeq 1$)
\begin{equation}
n=n_{\max }\frac{1+\mathrm{erf}[\alpha (t-t_{m})]}{2}
,  \label{ngauss}
\end{equation}
with
\begin{equation}
n_{\max }={\sqrt{\pi }\Gamma _{m}}/{\alpha }.
\label{nmax}
\end{equation}
The function $(1+\mathrm{erf})/2$ varies between $0$ and $1$, hence $n_{\max
}$ is the maximum density of bubbles that can be nucleated. For
$n_{\max}H^{-3}<1$ there will always be, in average, less than a bubble per
Hubble volume, and the time $t_N$ will not exist. Hence, the existence of
$t_N$ gives the condition
\begin{equation}
\frac{\Gamma _{m}}{H^{4}}>\frac{1}{\sqrt{\pi }}\frac{\alpha }{H}.
\label{condnb1}
\end{equation}
If this condition is fulfilled, then according to Eq.~(\ref{ngauss}) a
number of bubbles $\sim \Gamma_m/(H^3\alpha)$ will nucleate in a Hubble
volume in a time interval of order $\alpha ^{-1}$ around $t=t_{m}$. On the
other hand, if this condition is not fulfilled one expects that the phase
transition will be stuck in the false vacuum.

\begin{figure}[bt]
\centering
\epsfysize=5cm \leavevmode \epsfbox{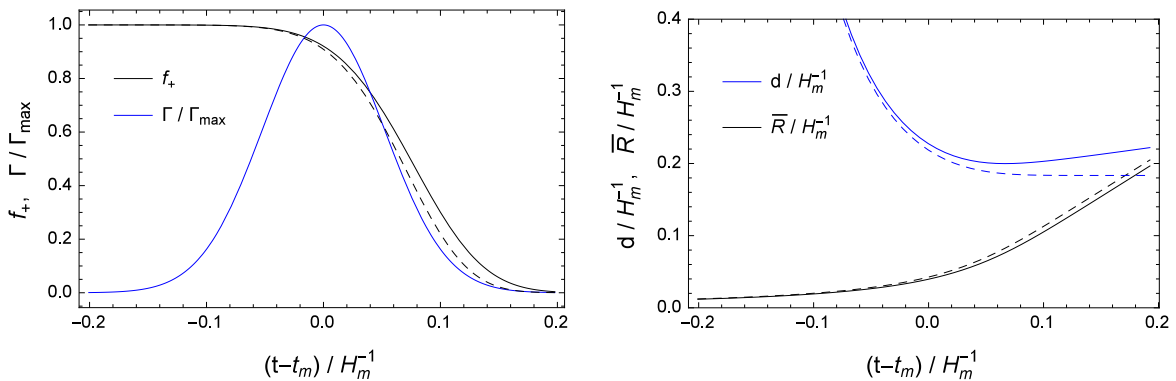}
\caption{The evolution of $f_{+}$, $\Gamma $, $d$ and $\bar{R}$ for a gaussian
nucleation rate, for $v_{w}=1$, $\protect\alpha /H=13.5$ and $\Gamma
_{m}/H^{4}=1505$.  Dashed lines are
obtained from Eq.~(\ref{Igauss}) while solid lines are obtained from Eq.~(\ref{iga}).}
\label{figapcuad}
\end{figure}

The fraction of volume is given by $f_{+}=\exp (-I)$, with $I$ given in this
case by
\begin{equation}
I(t)=I_m\left[(1+x^{2})e^{-x^{2}}+\sqrt{\pi}(3x+2x^{3})\frac{1+
\mathrm{erf}(x)}{2}\right],   \label{Igauss}
\end{equation}
where $x=\alpha(t-t_m)$ and
\begin{equation}
I_{m}=\frac{2\pi v_{w}^{3}\Gamma _{m}}{3\alpha ^{4}} .  \label{Im}
\end{equation}
For $t\lesssim t_{m}-\alpha ^{-1}$ we have $ I(t)\simeq 0$, while at $t=t_m$
we have $I=I_m$ and for $t\gtrsim t_{m}+\alpha ^{-1}$ we have $I(t)\simeq
\sqrt{\pi }I_{m}\alpha (t-t_{m})[3+2\alpha ^{2}(t-t_{m})^{2}]$. This cubic
growth contrasts with the exponential behavior in Eq.~(\ref{Iexp}). In
Fig.~\ref{figapcuad} we show the evolution of some quantities for $v_{w}=1$,
$\alpha /H\simeq 13.5$ and $\Gamma _{m}/H^{4}\simeq 1505$, which are the
values corresponding to the physical case considered in the third row of
Fig.~\ref{figevol} (the time and distance scales in the two figures do not
coincide due to the different normalization factors $H_{c}$ and $H_{m}$; we
compare with the physical model in more detail below).
The dashed lines in Fig.~\ref{figapcuad} are obtained from
Eq.~(\ref{Igauss}) while the solid lines take into account the scale factor
(see below). Although the latter reproduce better the behavior of the curves
in Fig.~\ref{figevol}, we see that the quantitative difference between the
two approximations is not significant.

\begin{figure}[bt]
\centering
\epsfysize=7cm \leavevmode \epsfbox{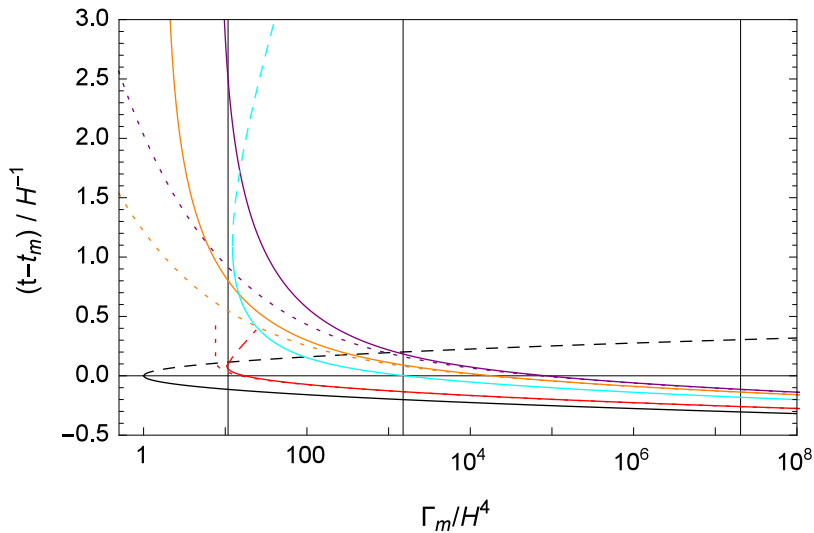}
\caption{Some reference times for the gaussian approximation, for $v_{w}=1$ and
$\protect\alpha /H=13.5$. Solid and dashed black lines:
the times at which $\Gamma=H^4$. Solid and dashed red lines: the times at which $nH^{-3}=1$.
Solid orange and purple lines: the times at which $f_+=e^{-1}$ and $f_+=0.01$, respectively.
Dotted lines correspond to neglecting the variation of the scale factor. The cyan lines indicate
the times at which $dI/dt=3H$.
The vertical lines indicate the values $\Gamma_m/H^4\simeq 2\times 10^7$, $1505$ and $11$,
corresponding to the cases $A/v=0.126$, $0.128$ and
$0.129$ of Fig.~\ref{figpoints}, respectively.}
\label{figtimes}
\end{figure}

The curves in Fig.~\ref{figapcuad} are plotted from $t=t_{H}$, which is
readily obtained from the nucleation rate (\ref{gammagauss}),
\begin{equation}
t_{H}-t_{m}=-\alpha ^{-1}\sqrt{\log (\Gamma _{m}/H^{4})}.
\label{tHg}
\end{equation}
This time is shown in Fig.~\ref{figtimes} (black solid line). The black
dashed line corresponds to the time at which $\Gamma$ reaches the value
$H^4$ again, after crossing its maximum, and is obtained by changing the
sign in the right hand side of Eq.~(\ref{tHg}). This time gives an
indication of when  $\Gamma$ becomes small again. Notice that $\log(\Gamma
_{m}/H^4)\simeq S_H-S_m$. Hence, if the phase transition takes place around
the minimum of $S$, we will have $t_{m}-t_{H}=$ a few $ \alpha ^{-1}$. This
can be appreciated in Fig.~\ref{figtimes} (where
$\alpha^{-1}\sim0.1H^{-1}$).

The time $t_{N}$ is obtained from Eq.~(\ref{ngauss}), and is given by the
inverse of the error function,
\begin{equation}
t_{N}-t_{m}=\alpha ^{-1}\mathrm{erf}^{-1}\left[ \frac{2}{\sqrt{\pi }}\frac{
\alpha /H}{\Gamma _{m}/H^{4}}-1\right] .
\label{tng}
\end{equation}
This time is indicated by a red dotted line in Fig.~\ref{figtimes}, which
coincides with the red solid line in most of the range. The latter takes
into account the variation of the scale factor (see below). The two
approximations only separate for $\Gamma_m/H^4$ very close to the limit in
which $t_N$ ceases to exist. Even this lower bound is very similar for the
two approximations. If we ignore the scale factor, this limit is given by
the condition (\ref{condnb1}), where the curve of the function
$\mathrm{erf}^{-1}$ becomes vertical. For smaller values of $\Gamma_m/H^4$
the final number of bubbles nucleated in a Hubble volume is less than 1.

The times $t_{I}$, $t_{P}$, $t_{E}$, and $t_{F}$, which correspond to
specific values of $I$, are obtained by inverting Eq.~(\ref{Igauss}). This
gives time intervals of the form $t_i-t_m=\alpha^{-1}x_i$, where $x_i$ is a
dimensionless function of $\alpha^4/\Gamma_m$. In Fig.~\ref{figtimes} we
show the times $t_E-t_m$ (orange) and $t_F-t_m$ (purple). The curves with
dotted lines are those obtained from Eq.~(\ref{Igauss}), while the curves
with solid lines take into account the variation of the scale factor and are
given by Eq.~(\ref{iga}) below. We see that ignoring the scale factor is a
good approximation for $\Gamma_m/H^4\gtrsim 10^2$ ($S_H-S_m\gtrsim 5$), in
which case we have $|t_i-t_m|\sim\alpha^{-1}$. This applies, for instance,
to the values $\Gamma_m/H^4\simeq 1505$ and $\Gamma_m/H^4\simeq 2\times
10^{7}$ (indicated by vertical lines), which correspond to the
cases\footnote{Actually, the value $\alpha/H\simeq 13.5$ used in
Fig.~\ref{figtimes} corresponds only to the case $A/v=0.128$. Nevertheless,
since $\alpha$ has little variation, Fig.~\ref{figtimes} gives a good
qualitative, and even quantitative, description of the other cases as well.}
$A/v=0.128$ and $A/v=0.126$ considered in Sec.~\ref{dyn}, respectively. On
the other hand, for small values of $\Gamma_m/H^4$ the duration of the phase
transition becomes of order $H^{-1}$ and the variation of the scale factor
becomes relevant.

The  average nucleation rate $\bar\Gamma(t)=\Gamma(t) \exp[-I(t)]$ is an
important quantity since it determines the bubble number density as well as
the bubble size distribution and, hence, the various size scales. Obtaining,
e.g., the bubble separation or the average radius from Eq.~(\ref{Igauss})
requires a numerical computation. Nevertheless, assuming that the nucleation
rate turns off before $f_+$ decreases, we may just set $\bar
\Gamma(t)=\Gamma(t)$. This rough approximation is actually quite good in the
case of Fig.~\ref{figapcuad} (as can be seen also in the left bottom panel
of Fig.~\ref{figevol}), and it is even better for stronger phase
transitions. In this approximation all the bubbles nucleate for $f_+\simeq
1$, and the number density is given by Eq.~(\ref{ngauss}). Thus, the average
bubble separation achieves a final value $d\simeq n_{\max}^{-1/3}\simeq
\pi^{-1/6}(\alpha/\Gamma_m)^{1/3}$. On the other hand, the radius
distribution is  given by
\begin{equation}
\frac{dn}{dR}(t)\simeq v_{w}^{-1}\,\Gamma_m\exp[-\alpha^2(t-t_m-R/v_{w})^2].
\label{dndRg}
\end{equation}
This distribution  has a width $\Delta R\sim v_w\alpha^{-1}$ and its peak is
given by
\begin{equation}
R_p\simeq v_w(t-t_m)
\label{rpgap}
\end{equation}
(which is negative for $t<t_m$). On the other hand, the peak of the
volume-weighted distribution is given by
\begin{equation}
R_p'\simeq\frac{v_w}{2}[t-t_m+\sqrt{(t-t_m)^2+6\alpha^{-2}}].
\label{rpprigap}
\end{equation}
Therefore, at $t=t_m$ we already have a positive peak $R_p'\simeq
1.2v_w\alpha^{-1}$. This gives a good description of the behavior of the
peaks in the right panels of Figs.~\ref{figdistrib} and \ref{figndistrib}.
As $t$ increases $R_p$ and $R_p'$ approach each other, since for
$t-t_m\gg\alpha^{-1}$ we have $R_p'\simeq v_w(t-t_m)$. This behavior can
also be appreciated in Fig.~\ref{figndistrib}.

\subsection{Gaussian nucleation rate with exponential expansion}

If we take into account the variation  of the scale factor, we may still
obtain analytic expressions, as long as we assume a simple form for $a(t)$
or, equivalently, for $H=\dot a/a$. For  a short enough time $\Delta t\ll
H^{-1}$ this is not relevant, and we may set $H(t)\simeq H(t_m)$.  On the
other hand,  longer times $\Delta t\gtrsim H^{-1}$ only matter in cases in
which there is  strong supercooling. In such a case the energy density
begins to be dominated by the false vacuum (since $\rho_R/\rho_V\sim
T^4/v^4$), and the expansion rate will again be approximately constant.
Therefore, a constant $H$ should be a good approximation in general for very
strong phase transitions, and we have
\begin{equation}
a(t')/a(t)=\exp[H(t'-t)].
\label{aexp}
\end{equation}
In Ref.~\cite{tww92}, such an exponential expansion was considered with an
exponential nucleation rate. However, as we have seen, for the cases in
which the exponential rate is a good approximation, the duration of the
phase transition is generally short enough, so that the scale factor can be
safely ignored, i.e., $\Delta t\sim\beta^{-1}\ll H^{-1}$. In contrast, if
the minimum of $S$ is reached we may have $\Delta t>\alpha^{-1}$.

We thus  use the relation (\ref{aexp}) together with  the gaussian
nucleation rate (\ref{gammagauss}) in Eqs.~(\ref{radio})-(\ref{dndR}). The
initial number density of bubbles (i.e., for $f_+\simeq 1$) is given by
\begin{equation}
n=n_{\max }\frac{1+\mathrm{erf}[\alpha (t-t_{m})]}{2}
\exp\left[-3H(t-t_m)+\left(\frac{3}{2}\frac{H}{\alpha}\right)^2\right],
\label{nga}
\end{equation}
with $n_{\max}$ defined in Eq.~(\ref{nmax}). This result is similar to
Eq.~(\ref{ngauss}), with the additional exponential factor which takes into
account the dilution $n\sim a^{-3}$. For $t-t_m\sim\alpha^{-1}$, we have in
general $H(t-t_m)\sim H/\alpha\ll 1$ and we recover Eq.~(\ref{ngauss}). Due
to this suppression, the maximum number of bubbles which can be nucleated in
a Hubble volume is now smaller than $n_{\max}H^{-3}$. As a consequence, we
have a stronger condition than Eq.~(\ref{nmax}) for the existence of the
time $t_N$. Nevertheless, for the typical values $H/\alpha\sim 0.1$ the
difference is small, as can be seen in Fig.~\ref{figtimes} (solid and dotted
red lines), and the bound on $\Gamma_m/H^4$ is still roughly given by
Eq.~(\ref{condnb1}). Thus, according to Eq.~(\ref{nga}), a total bubble
density $n\simeq n_{\max}$ is nucleated in a time $t-t_m\sim\alpha^{-1}$,
and for $t-t_m\gtrsim \alpha^{-1}$  the error function saturates and we have
(neglecting the term $H^2/\alpha^2$)
\begin{equation}
nH^{-3}\simeq n_{\max}H^{-3}e^{-3H(t-t_m)}. \label{ngaap}
\end{equation}
For $t-t_m\sim H^{-1}$ the exponential suppression becomes important, and
eventually the condition $nH^{-3}=1$ will be achieved again (red dashed line
in Fig.~\ref{figtimes}). This second solution only makes sense if it is
achieved during the phase transition; once the universe is in the true
vacuum the assumption of an exponential expansion will not be a good
approximation.

For a constant wall velocity, Eq.~(\ref{radio}) gives
$R(t',t)=v_wH^{-1}(e^{H(t-t')}-1)$. Then Eq.~(\ref{I}) gives, defining
$\tau=H(t-t_m)$,
\begin{eqnarray}
I(t) &=& I_{\max}\left[\frac{1+\mathrm{erf}(\frac{\alpha}{H}\tau)}{2}
-3\,e^{-\tau+\frac{H^2}{4\alpha^2}}\,\frac{1+\mathrm{erf}[\frac{\alpha}{H}
(\tau-\frac{1}{2}\frac{H^2}{\alpha^2})]}{2}
\right. \label{iga} \\ & & \left.
 +\,3\,e^{-2\tau+\frac{H^2}{\alpha^2}}\,\frac{1+\mathrm{erf}
 [\frac{\alpha}{H}(\tau-\frac{H^2}{\alpha^2})]}{2}
-\,e^{-3\tau+\frac{9H^2}{4\alpha^2}}\frac{1+\mathrm{erf}[\frac{\alpha}{H}
(\tau-\frac{3}{2}\frac{H^2}{\alpha^2})]}{2}
\right], \nonumber
\end{eqnarray}
where
\begin{equation}
I_{\max}=\frac{4\pi^{3/2} v_w^3\Gamma_m}{3H^3\alpha}.
\end{equation}
The result (\ref{Igauss}) is recovered if we assume
$t-t_m\lesssim\alpha^{-1}$ and expand Eq.~(\ref{iga}) to third order in
$H/\alpha$. On the other hand, for $t-t_m\gtrsim\alpha^{-1}$ the expression
becomes even simpler, since the error functions saturate and we obtain
(neglecting the terms $H^2/\alpha^2$),
\begin{equation}
I(t)\simeq I_{\max}\left(1-e^{-H(t-t_m)}\right)^3. \label{Iaap}
\end{equation}
A numerical comparison of Eqs.~(\ref{Igauss}) and (\ref{iga}) is given by
the orange and purple lines in Fig.~\ref{figtimes}, corresponding to the
times at which the values $I=1$ ($t=t_E$) and $I=\log 100$ ($t=t_F$) are
reached, respectively. The solid lines are obtained from Eq.~(\ref{iga}),
while the dotted lines ignore the variation of the scale factor. The latter
is a very good approximation for $t-t_m\lesssim \alpha^{-1}$. On the other
hand, for $t-t_m\gtrsim \alpha^{-1}$ we may use the approximation
(\ref{Iaap}), and the equation $I(t_i)=I_i$ gives
\begin{equation}
t_i-t_m\simeq H^{-1} \log\left[\frac{1}{1-(I_i/I_{\max})^{1/3}}\right].
\label{titmga}
\end{equation}
For $\alpha/H\sim 10$, this becomes a very good approximation for $t-t_m> 2
\alpha^{-1}$ ($\Gamma_m< 0.02I_i\alpha^4$). In the range
$\alpha^{-1}\lesssim t-t_m\lesssim H^{-1}$ we may use the approximation
$I_i/I_{\max}\ll 1$ in Eq.~(\ref{titmga}), which gives
\begin{equation}
t_i-t_m\simeq \frac{1}{\sqrt{\pi}v_w}\left(\frac{3 I_i}{4}\frac{\alpha}{\Gamma_m}\right)^{1/3}.
\label{titmga2}
\end{equation}

As can be seen by inspection of Eq.~(\ref{iga}), and more directly in the
approximation (\ref{Iaap}), for $t-t_m\gg H^{-1}$ we have a constant
$I(t)=I_{\max}$. Formally, this implies that the phase transition never
ends, since $f_+(t)$ approaches the asymptotic value $e^{-I_{\max}}>0$.
Nevertheless, notice that even when $I(t)$ grows exponentially, as in
Eq.~(\ref{Iexp}), the limit $f_+=0$ is  reached in an infinite time. This is
a characteristic of the approximation (\ref{fmas}-\ref{I}), which lead us to
define the time $t_F$ by $f_+(t_F)=0.01$ in Sec.~\ref{dyn}. In the present
case, for large enough $I_{\max}$ the fraction of volume in the false vacuum
will reach a very small value within a Hubble time, and this means that the
phase transition is actually completed. On the other hand, for smaller
values of $I_{\max}$ the situation changes. Reaching the value $I=I_F$
requires $I_{\max}>I_F$, which gives the restriction
\begin{equation}
\frac{\Gamma_m}{H^4}>\frac{3}{4\pi^{3/2}}\frac{I_F}{v_w^{3}}\frac{\alpha}{H}.
\label{asympt}
\end{equation}
As  this bound is approached, the time needed to reach the value $I=I_F$
becomes infinite (and we have a similar bound for any value $I=I_i$). This
is apparent in Eq.~(\ref{titmga}) and can be appreciated in
Fig.~\ref{figtimes}, since the purple  solid line has an asymptote at this
value of ${\Gamma_m}/{H^4}$ (and similarly for the orange solid line).

It is worth mentioning that, in contrast to the gaussian rate, an
exponential nucleation rate $\Gamma\propto e^{\beta(t-t_*)}$ gives an
exponentially growing $I(t)\propto\Gamma(t)$, even in the case of an
exponential scale factor. In spite of this, for small $\beta$ the completion
of the transition will take many Hubble times (see Ref.~\cite{tww92} for
details). In this scenario, it is convenient to consider the physical volume
in the false vacuum, $V_{\mathrm{phys}}(t)\propto a^3(t)f_+(t)$. This volume
always grows at the beginning of the phase transition, and in the case of
small $\beta$, it will grow during most of the phase transition. Hence,
there is a new ``milestone'', namely, the time $t_e$ at which the derivative
\begin{equation}
{dV_{\mathrm{phys}}}/{dt}=\left(3H-{dI}/{dt}\right){V_{\mathrm{phys}}}
\end{equation}
vanishes. For the exponential nucleation the time $t_e$ always exists, but
for $\beta\lesssim H$ this time is later than $t_F$ \cite{tww92}. In such a
case, at $t=t_F$ the physical volume in the false vacuum has never stopped
growing, and we cannot say that the phase transition is coming to an end.
Nevertheless, as we have seen, in general we have $\beta\gg H$.

For the gaussian nucleation rate, even though we have in general
$\alpha^{-1}\ll H^{-1}$,  the phase transition can take a longer time. For
$t-t_m\gtrsim \alpha^{-1}$ we may use the approximation (\ref{Iaap}), and we
see that $dI/dt$ reaches a maximum at $(t-t_m)=\log(3)H^{-1}$. The maximum
value is $(4/9)HI_{\max}$, so the existence of the time $t_e$ requires
$I_{\max}>27/4$, which gives another bound on $\Gamma_m$,
\begin{equation}
\frac{\Gamma_m}{H^4}>\frac{3}{4\pi^{3/2}}\frac{27/4}{v_w^{3}}\frac{\alpha}{H}.
\label{condte}
\end{equation}
For smaller $\Gamma_m$, the volume in the false vacuum always grows and the
phase transition never completes. Notice that this bound is slightly more
restrictive than Eq.~(\ref{asympt}) (since $27/4>I_F$). For larger
$\Gamma_m$ the time $t_e$ exists, and it can be checked that it is earlier
than $t_F$. Therefore, in this case $dV_{\mathrm{phys}}/dt$ becomes negative
before $t=t_F$. However, since $dI/dt$ decreases for $t-t_m>\log (3)H^{-1}$,
there is a second time $t_e'$ at which $dV_{\mathrm{phys}}/dt$ becomes
positive again. Requiring that this does not happen before the time $t_F$
gives the condition
\begin{equation}
\frac{\Gamma_m}{H^4}>\frac{3(I_F+1)^3}{4\pi^{3/2}v_w^{3}I_F^2}\frac{\alpha}{H},
\label{bound}
\end{equation}
which is a little more restrictive than (\ref{asympt}) and (\ref{condte}).
In Fig.~\ref{figtimes} the times $t_e$ and $t_e'$ are indicated by solid and
dashed cyan lines, respectively. We shall take the condition (\ref{bound})
as the bound for the completion of the phase transition. For larger
$\Gamma_m/H^4$, when the value $f_+=0.01$ is reached the physical volume in
the false vacuum is  decreasing, and the phase transition will soon
complete.

For the case of Fig.~\ref{figtimes} ($\alpha/H\simeq 13.5$) the bounds given
by Eqs.~(\ref{asympt}), (\ref{condte}), and (\ref{bound}) are
$\Gamma_m/H^4\simeq 8.4$, $12.3$, and $15.1$, respectively. The leftmost
vertical line in this figure lies between the first two values. For this
case we cannot say that the phase transition will ever complete, even though
the value $f_+=0.01$ is reached. We also see that this value of $f_+$ is
achieved in a relatively long time, $t_F-t_H\simeq 2.5H^{-1}$. This case
($\Gamma_m/H^4\simeq 11$) corresponds to the case $A/v=0.129$ of the
physical model, and was considered in Fig.~\ref{figpoints} (second curve
from the top). In that figure, the long duration of the phase transition is
apparent from the low temperatures at which the points $P$ and $E$ are
reached (the value $T_F\simeq 0.1$ lies outside the range of the figure).
Notice also that in Fig.~\ref{figtimes} the vertical line barely touches the
solid red line. Accordingly, in the numerical computation of Sec.~\ref{dyn}
the value  $nH^{-3}=1$ is exceeded only during a short time (the maximum
number of bubbles achieved is $nH^{-3}\simeq 1.2$).

To estimate the bubble size distribution we must now take into account the
scale factors in Eq.~(\ref{dndR}). At a given time $t$, the time $t_R$ at
which a bubble of radius $R$ was nucleated is given by
$t_R=t-H^{-1}\log(HR/v_w+1)$, and the scale factors at these times are
related by $a/a_R=1+HR/v_w$. As in the previous subsection, we shall use the
approximation $\bar\Gamma(t)\simeq\Gamma(t)$, which is equivalent to
assuming that $f_+(t_R)\simeq 1$ at the time in which most bubbles were
nucleated. Thus, the size distribution is given by
\begin{equation}
\frac{dn}{dR}(t)=\frac{\Gamma_m\exp\left\{-\alpha^2
\left[t-t_m-H^{-1}\log(1+HR/v_w)\right]^2
\right\}}{v_w(1+HR/v_w)^4} \label{dndRga}
\end{equation}
This reduces to Eq.~(\ref{dndRg}) for small $HR$. The main difference is in
the denominator of Eq.~(\ref{dndRga}), which takes into account the dilution
of the nucleated bubbles. Thus, for increasing time the peak,
$R_p=v_wH^{-1}(e^{H(t-t_m)}-1)$, moves to higher $R$, and the amplitude at
$R=R_p$ decreases as $(1+HR_p/v_w)^{-4}$. This is in agreement with the
right panel of Fig.~\ref{figdistrib}. We see that for large $t$ the radius
growth will be dominated by the exponential expansion, as any physical
length. This effect will be noticeable if the completion of the phase
transition takes longer than $H^{-1}$. For $t-t_m\gtrsim\alpha^{-1}$ but
still smaller than $H^{-1}$, we have $R_p\simeq v_w(t-t_m)[1+H(t-t_m)/2]$,
i.e., the result (\ref{rpgap}) has a small modification.

Similarly, for the peak of the volume-weighted distribution, to linear order
in $H(t-t_m)$ we obtain the result (\ref{rpprigap}). Using the approximation
(\ref{titmga2}), we have $R_p(t_F)\simeq R_p'(t_F)\simeq 0.85
(\alpha/\Gamma_m)^{1/3}[1+\mathcal{O}(H(t-t_m))]$. On the other hand, the
average bubble separation $d=n^{-1/3}$ reaches a minimum value $d\simeq
n_{\max}^{-1/3}\simeq \pi^{-1/6}(\alpha/\Gamma_m)^{1/3}$ in a time
$t-t_m\sim\alpha^{-1}$, and for $t-t_m\gtrsim\alpha^{-1}$ it grows,
according to Eq.~(\ref{ngaap}), as $d\simeq n_{\max}^{-1/3}e^{H(t-t_m)}$.
Therefore, for $t_F-t_m\lesssim H^{-1}$ we have
$d(t_F)\simeq\pi^{-1/6}(\alpha/\Gamma_m)^{1/3}[1+H(t_F-t_m)] \simeq
0.83(\alpha/\Gamma_m)^{1/3}[1+\mathcal{O}(H(t-t_m))]$, which is very similar
to the values of $R_p$ and $R_p'$.

\section{Application to specific models} \label{aplic}

\subsection{Computing the parameters}

In order to apply the analytical approximations obtained in the previous
section, it is only necessary to compute, for a given model, either the
quantities $\Gamma _{\ast },\beta ,v_{w}$ at a suitable temperature $T_*$ or
the quantities $\Gamma _{m},\alpha ,v_{w}$ at $T=T_m$. Since $T_{m}$ is just
the minimum of $S(T)$, its computation is relatively simple, as it does not
involve the dynamics of the phase transition. On the other hand, there is
some freedom in choosing the value of $T_{\ast }$. In principle, we might
take, e.g., any value between $T_{H}$ and $T_{F}$, provided that the
linearization of $S$ is valid in the whole range. Notice that $T_{\ast
}=T_{H}$ is the simplest choice, since $T_{H}$ is obtained by solving
Eq.~(\ref{tH}), which does not involve the dynamics of the phase transition
(and therefore is as simple as obtaining $T_{m}$). However, for a real model
the linear approximation for $ S $ is valid only in a small time interval,
and the parameter $\beta $ may vary significantly in the interval
$[t_{H},t_{F}]$ (see Table \ref{tabbeta}). Therefore, the computations will
be more precise if $\beta $ is evaluated at a later time, such as $ t_{\ast
}=t_{E}$, where most bubbles are nucleated. It is relatively easy to follow
the phase transition up to the time $t_{N}$, since it only requires to
evaluate the integral\footnote{The time-temperature relation is often
approximated by $t\propto M_P/T^2$, which is obtained by assuming
$\rho_+\propto T^4$ in Eq.~(\ref{adexp}), i.e., neglecting the vacuum energy
density.} in Eq.~(\ref{nb}) with $\bar\Gamma=\Gamma$. Continuing to later
times, instead, involves solving numerically Eqs.~(\ref{radio})-(\ref{I}),
which is precisely the calculation one is trying to avoid by using
approximations such as (\ref{gammaexplin}) or (\ref{gammagauss}). Therefore,
it is very common to evaluate $\beta$, as well as $v_w$, at $T=T_{N}$.

A relatively simple and accurate approximation for $T_N$ can be obtained by
using the exponential rate approximation with $t_*=t_N$. Although this will
not be a good approximation for computing the whole development of the phase
transition (which involves times  $t-t_N\sim 10\beta^{-1}$), we can use it
to compute the number of bubbles nucleated before $t_N$. Indeed, since
$\Gamma$ falls quickly for $t<t_N$, the integral (\ref{nb}) effectively
involves times $t_N-\beta^{-1}\lesssim t<t_N$. The result (assuming a
constant scale factor and wall velocity in the time $\sim\beta^{-1}$) is
given by Eq.~(\ref{gatnlin}), with $\beta $ evaluated at $t_{\ast }=t_{N}$.
In terms of $T_N$ we have
\begin{equation}
S(T_{N})-\frac{3}{2}\log \left[ \frac{S(T_{N})}{2\pi }\right] =4\log \left[
\frac{T_{N}}{H(T_{N})}\right] -\log [T_{N}S^{\prime }(T_{N})].  \label{TNap}
\end{equation}
The same reasoning can be applied for the computation of $I(t)$ (see, e.g.,
\cite{eikr92}). As long as the time $t$ is far enough from the minimum,
$t_m-t\gtrsim\alpha^{-1}$, the nucleation rate falls quickly for $t'<t$ in
the integral (\ref{I}). Then, we may use the exponential rate, and the best
choice for $t_*$ is $t_*=t$. For instance, to estimate the temperature
$T_{E}$ we may set $t=t_{\ast }=t_{E}$ in Eq.~(\ref{Iexp}), which gives
$8\pi v_{w}^{3}\Gamma (T_{E})/\beta ^{4}=1$, with $\beta$ evaluated at
$T_E$. We thus have the equation
\begin{equation}
S(T_{E})-\frac{3}{2}\log \left[ \frac{S(T_{E})}{2\pi }\right] =4\log \left[
\frac{T_{E}}{H(T_{E})}\right] -4\log [T_{E}S^{\prime }(T_{E})]+\log (8\pi
v_{w}^{3}).  \label{TEap}
\end{equation}
The wall velocity in the last term  can be ignored for very strong phase
transitions, i.e., setting $v_{w}=1$ will not introduce a significant error
in the value of $T_{E}$. These equations are very similar to that for
$T_{H}$, and give $T_N$ and $T_E$ without considering details of the
dynamics of the phase transition. The main difference with Eq.~(\ref{tH}) is
the appearance of a term involving the derivative of $S$, and the
approximations will break down if $S^{\prime }(T)$ becomes too small.
Nevertheless, they will provide good estimations for $T_N$ and $T_E$ quite
close to $T_m$. We have checked this by comparing with the values obtained
from the numerical evolution. For instance, in the very strong cases
$A/v=0.126$ and $A/v=0.128$, the difference in $T_N$ as given by
Eq.~(\ref{TNap}) or by the numerical evolution is on the order of a 0.1\%.
On the other hand, the agreement for $T_E$ is on the order of a $1\%$ in the
case $A/v=0.126$, while for $A/v=0.128$ the approximation (\ref{TEap})
breaks down since $S'$ becomes negative.

Some applications require to compute certain quantities at a given reference
temperature, for which solving Eq.~(\ref{TNap}) or (\ref{TEap}) may be
enough. For other applications the time intervals will also be important,
and the time-temperature relation may be needed.  Notice that estimations of
time intervals such as Eqs.~(\ref{tnexp}-\ref{tftiexp}), which assume a
constant $\beta$, cannot be used to determine the time elapsed from the
critical temperature. If the equation of state is simple enough, the
Friedmann equation will give an analytic expression for the time-temperature
relation. For instance, for the bag EOS the expansion rate is of the form
$H=\sqrt{a+bT^{4}}$. In this case Eq.~(\ref{adexp}) can be solved
analytically. Using the condition $T=\infty $ at $t=0$, we have \cite{kk86}
$T^{2}=\sqrt{a/b}/\sinh (2\sqrt{a}t)$. Our model (\ref{freeen}) has the bag
form in the $+$ phase\footnote{For a different model, the bag EOS can be
used as an approximation, just as we did for the $-$ phase of our model in
Sec.~\ref{dyn}.}, and we have $a=(8\pi /3)\rho _{V}/M_{P}^{2}$ and $b=(4\pi
^{3}g_{\ast }/45)M_{P}^{-2}$. Thus, we may write
\begin{equation}
\frac{T^{2}}{T_{c}^{2}}=\frac{\sqrt{\rho _{V}/\rho _{R}(T_{c})}}{\sinh (2
\sqrt{\rho _{V}/\rho (T_{c})}H_{c}t)} . \label{Tt}
\end{equation}
This equation is valid while the system is in the $+$ phase, and thus it can
be inverted to obtain the time $t_{c}$ as well as the time $t_{N}$.
Moreover, as already discussed, for a phase transition mediated by
detonations or runaway walls the average energy density is given by
$\rho_+(T)$ until the end of the transition. Therefore, Eq.~(\ref{Tt}) is
still valid at later times, and we have
\begin{equation}
\frac{t-t_{c}}{H_{c}^{-1}}=\frac{1}{2}\sqrt{\frac{\rho (T_{c})}{\rho _{V}}}
\left[ \sinh ^{-1}\left( \sqrt{\frac{\rho _{V}}{\rho _{R}(T_{c})}}\frac{
T_{c}^{2}}{T^{2}}\right) -\sinh ^{-1}\left( \sqrt{\frac{\rho _{V}}{\rho
_{R}(T_{c})}}\right) \right] .
\label{tT}
\end{equation}
We remark that Eq.~(\ref{tT}) is independent of the nucleation rate and can
be used, e.g., to compute the time $t_{m}$ if the temperature $T_m$ is
reached during the phase transition. After the phase transition there will
be some reheating, and Eq.~(\ref{Tt}) will no longer hold.

\subsection{Comparison with the numerical computation}

We shall now compare the different approximations with the numerical
computation. For that aim we shall consider the evolution of the fraction of
volume, as well as the normalized distribution of bubble sizes at the time
$t=t_E$. Thus, for instance, for the exponential rate we have
$f_+(t)=\exp[-I_*\exp(\beta(t-t_*))]$ and
\begin{equation}
\frac{1}{n}\frac{dn}{dR}({t_E})=\frac{\beta}{v_w}
\frac{\exp[-\beta R/v_w-e^{-\beta R/v_w}]}{1-e^{-1}}. \label{dndRexptE}
\end{equation}
Notice that the latter equation does not depend explicitly on $t_*$.
Different choices of this time will be reflected in Eq.~(\ref{dndRexptE})
through different values of $\beta$ and $v_w$. The parameter $I_*$ appearing
in $f_+$ is also a function of $\beta$ and $v_w$. For instance, for
$t_*=t_N$ we have $I_*=8\pi v_w^3(H/\beta)^3$, and for $t_*=t_E$ we just
have $I_*=1$. On the other hand, the value of $t_*$ is related to the
corresponding temperature $T_*$ by the analytic expression (\ref{tT}). We
shall consider the temperatures $T_*=T_{N}$ and $T_*=T_E$, which can be
obtained from the semi-analytic approximations (\ref{TNap}) and
(\ref{TEap}), respectively. We remark that, in these approximations, the
only parameters which we shall compute numerically are $T_*$, $\beta(T_*)$
and $v_w(T_*)$. Similarly, for the gaussian rate the results depend only on
the numerical values of $\alpha(T_m)$ and $\Gamma(T_m)$. The function
$f_+(t)$ depends on the parameter $t_m$, which is again obtained from
Eq.~(\ref{tT}), while the size distribution depends only on the difference
$t_E-t_m$, which is obtained from Eq.~(\ref{Igauss}) as a function of
$\alpha$ and $\Gamma_m$.

\begin{figure}[tbp]
\centering
\epsfxsize=15cm \leavevmode \epsfbox{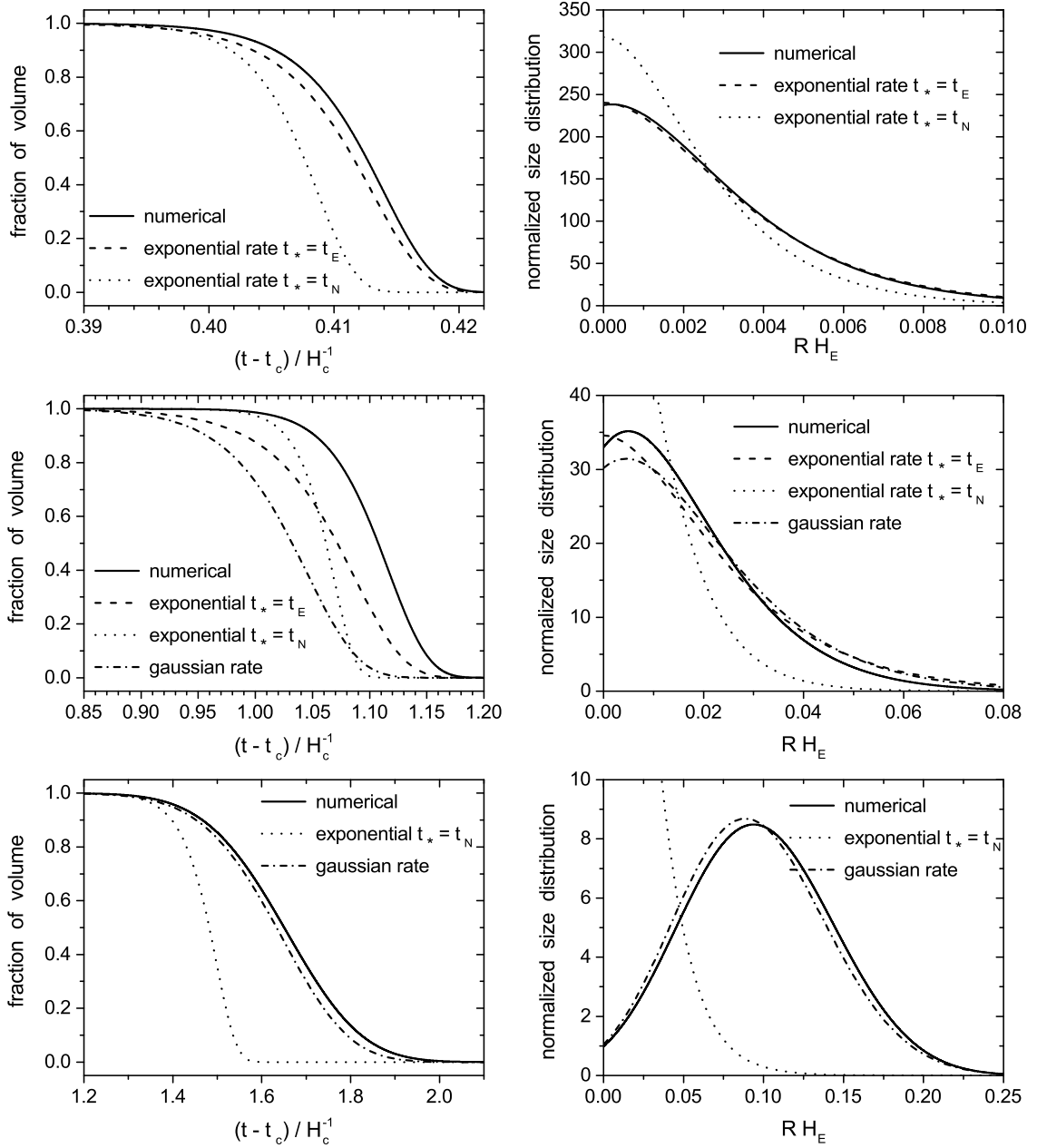}
\caption{Comparison of the evolution of
the fraction of volume (left panels) and the normalized radius distribution at $t=t_E$
(right panels), for $A/v=0.11$ (top), $A/v=0.126$ (middle) and
$A/v=0.128$ (bottom).}
\label{figcompara}
\end{figure}

In  Fig.~\ref{figcompara} we consider the cases discussed in Sec.~\ref{dyn}.
The solid lines correspond to the numerical computation. In the top panels
we consider the case $A/v=0.11$, and we show the results of the exponential
rate, with $t_*=t_E$ (dashed lines) and $t_*=t_N$ (dotted lines). As
expected, using $t_*=t_E$ gives an excellent approximation, while for
$t_*=t_N$ the approximation is not so good. We do not include the gaussian
approximation, since this case is far from the minimum of $S$.

The central panels of Fig.~\ref{figcompara} correspond to the case
$A/v=0.126$.  In the left panel it can be appreciated that, at the beginning
of the phase transition, the evolution of $f_+$ is better approximated by
the curve corresponding to $t_*=t_N$, while, as expected, that corresponding
to $t_*=t_E$ is closer to the numerical result in the last stages. On the
other hand, the gaussian rate  (dashed-dotted line) gives also a reasonably
good approximation for $f_+(t)$. Moreover, the shape of this curve describes
the evolution of the fraction of volume better than the curve of $t_*=t_N$.
Indeed, for the latter the evolution is too quick due to a large value of
$\beta(T_N)$. In the right panel it can be seen that the case $t_*=t_E$ and
the gaussian rate give quantitatively similar results for the size
distribution. However, the latter reproduces more correctly the shape of the
curve, with a non-vanishing maximum. Notice that this is a borderline case;
for lower values of $A/v$ the exponential rate gives a better approximation,
while for higher values the gaussian approximation will be better. As can be
seen in the figure, in this limit the error of both approximations is
smaller than a 10\%.

The  plots at the bottom of Fig.~\ref{figcompara} correspond to the case
$A/v=0.128$. In this case the gaussian rate is clearly a good approximation,
while the exponential approximation breaks down. Indeed, for $t_*=t_E$ we
have $\beta<0$ and the approximation cannot even be used. On the other hand,
we see in the left panel that the case $t_*=t_N$ is a good approximation
only for the first stages of the evolution.

\subsection{Implications for cosmic remnants} \label{implic}

As discussed in Sec.~\ref{conseq}, numerical simulations for computing
certain cosmological consequences of the phase transition require several
simplifications on the dynamics. One usual approximation is to assume a
constant wall velocity. In the case of a very strong phase transition this
is generally a good approximation since we have $v_w\simeq 1$. On the other
hand, in simulations bubbles are often nucleated simultaneously, which
corresponds to a delta-function nucleation rate, or at random times,
corresponding to a constant nucleation rate. The most realistic
approximation is an exponential rate of the form (\ref{gammaexplin}),
whereby the value of $\Gamma _{\ast }$ is generally chosen so that the
desired number of bubbles is nucleated in a sample volume. This number is
restricted by the limitations of the simulation. Nevertheless, the dynamics
of the phase transition is determined by the parameter $\beta$ rather than
by $\Gamma_*$. Thus, the results of the simulation will generally be
functions of $\beta $, as in Eq.~(\ref{GWcol}). As we have seen, this
exponential approximation for $\Gamma$ breaks down around the minimum of
$S(T)$, and the gaussian approximation is more appropriate. Our numerical
and analytic results show that several aspects of the dynamics, such as,
e.g., the bubble size distribution, cannot be directly extrapolated from
weaker phase transitions to very strong cases. To our knowledge, a gaussian
nucleation rate has not been considered so far in numerical simulations (and
it is certainly out of the scope of this work). In the gaussian case, the
dynamics will depend not only on the parameter $\alpha$ but also on
$\Gamma_m$.

In principle, one expects that the cosmic remnants depend particularly on
the bubble size distribution and, hence, on the nucleation rate. For
instance, the GWs inherit the wavenumber spectrum of the source
\cite{cds06}. However, recent simulations in the envelope approximation
\cite{w16} seem to contradict this expectation. In Ref.~\cite{w16} bubbles
were nucleated either using the exponential rate or simultaneously, and the
resulting GW spectrum is described in both cases by the same power laws at
high and low frequencies. Besides the similarity in the shape of the
spectra, the peak frequency and amplitude agree within the order of
magnitude, although in the exponential-rate case the frequency is lower and
the amplitude higher (see Fig.~3 in \cite{w16}). This similarity is somewhat
unexpected. Notice in particular that for a simultaneous nucleation the
bubble size distribution will peak around the average bubble separation,
while for an exponential rate smaller bubbles will be much more
abundant\footnote{Notice, though, that in Ref.~\cite{w16} the bubbles in the
simultaneous case were nucleated in the same positions as in the exponential
case, and therefore are not randomly distributed in space.}. In any case, we
remark that for GW generation one expects the \emph{volume-weighted} bubble
distribution to be more relevant. For the exponential rate, the peak of the
volume-weighted distribution has a value $R'_p\simeq 3.49 \,v_w/\beta$ at
the end of the transition, which is not too different from the final bubble
separation, $d\simeq 2.93\, v_w/\beta$. This may explain the similarities.
Moreover, if we assume that $R_p'$  is the relevant length in the
exponential nucleation while $d$ is the relevant length in the simultaneous
nucleation, their numerical difference is in agreement with the difference
in the peaks of the corresponding GW spectra. Notice also that even assuming
a direct relation $\tilde f = 1/R_p'$ gives (for $v_w\simeq 1$) $\tilde
f=0.29\,\beta$, which roughly agrees with the result of the envelope
approximation, Eq.~(\ref{GWcol}) above.

Regarding GWs generated by sound waves, the actual free parameter in the
simulations of Ref.~\cite{hhrw} is the average separation $d$, so this is in
principle the relevant quantity. On the other hand, the peak frequency in
Eq.~(\ref{GWsw}) suggests a relevant length scale $\simeq 0.3d$. In this
case, though, the width of the sound shell plays a relevant role \cite{h16}.
In any case, the results of the simulations of Ref.~\cite{hhrw} must be
interpreted with care, since relatively weak phase transitions were
considered [corresponding to the potential (\ref{potef}) with $A=0$], most
of them with deflagration walls, and it is not clear whether these results
extrapolate directly to stronger phase transitions. Beyond that, there is no
obstacle in formally extrapolating the approximations of Eq.~(\ref{GWsw}),
since $d$ is a well-defined physical quantity. In contrast, the
approximations (\ref{GWcol}) for the collision mechanism depend on the
parameter $\beta$ which becomes negative for strong phase transitions.

\begin{figure}[tb]
\centering
\epsfxsize=16cm \leavevmode \epsfbox{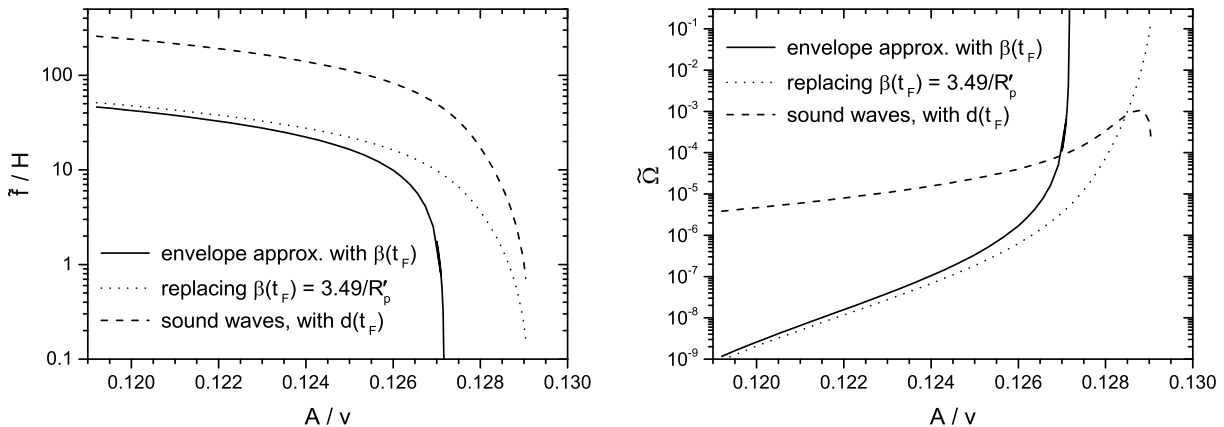}
\caption{Peak frequency and amplitude of gravitational waves at the end of the phase transition.}
\label{figgw}
\end{figure}

To illustrate this problem, we will apply Eqs.~(\ref{GWcol}) and
(\ref{GWsw}) to the model (\ref{potef}). We only consider the runaway
regime. For a runaway wall the quantities $\rho_{\mathrm{wall}}$ and
$\rho_{\mathrm{fl}}$ appearing in Eqs.~(\ref{GWcol}) and (\ref{GWsw}) are
given by \cite{urwalls}
\begin{equation}
\rho_{\mathrm{wall}}=F_{\mathrm{net}}^{UR},\qquad\rho_{\mathrm{fl}}\simeq
4(\rho_R+3\Delta\varepsilon-3F_{\mathrm{net}}^{UR})
(0.15v_-^2-0.132v_-^3+0.065v_-^4),
\end{equation}
where the quantities $F_{\mathrm{net}}^{UR}$, $\rho_R$, $\Delta\varepsilon$
and $v_-$ are defined in Sec.~\ref{dyn}. We evaluated these quantities, as
well as $d$ and $\beta$, at $t=t_F$. The result is shown in
Fig.~\ref{figgw}.
The solid lines correspond to the GWs generated by bubble collisions and the
dashed lines correspond to the result from sound waves. The latter have in
general a higher intensity, except for very strong phase transitions. This
is expected, since for stronger supercooling more energy goes to the bubble
walls and less energy goes to the fluid. This is why the GW intensity from
sound waves suddenly falls near the value $A/v=0.129$. For this model, this
is approximately the limit beyond which the phase transition is stuck in the
false vacuum while the temperature of the plasma decreases. On the other
hand, it is clear that the abrupt growth in $\tilde \Omega_{\mathrm{coll}}$
and the abrupt decrease in $\tilde f_{\mathrm{coll}}$ are due to the
vanishing of $\beta$ for $A/v\simeq 0.127$. Beyond this value, $\beta$
becomes negative and we cannot apply the approximation (\ref{GWcol}).

In order to extrapolate the results of Ref.~\cite{hk08}, where the
exponential-rate approximation was used, to very strong phase transitions,
we may first write Eqs.~(\ref{GWcol}) in terms of a physical length.
Assuming that the actual relevant scale in the collision mechanism is given
by the peak of the volume-weighted bubble distribution, as discussed above,
we set $\beta=3.49/R'_p(t_F)$, and Eqs.~(\ref{GWcol}) become
\begin{equation}
\tilde{f}_{\mathrm{coll} }\simeq 0.8 /R_p', \;\;
 \tilde\Omega _{\mathrm{coll}}\simeq 6.6\times 10^{-3}(\rho_{\mathrm{wall}}
 /{\rho _{\mathrm{tot}}})^{2}(HR'_p)^{2}.
\end{equation}
The result is plotted with a dotted line in Fig.~\ref{figgw}. We have also
considered the replacement $\beta=(8\pi)^{1/3}/d(t_F)$, which gives a very
similar result. We see that the solid and dotted lines begin to separate for
$A/v\simeq 0.126$, which for this model is the intermediate value between
the ranges of validity of the exponential and gaussian rates. Beyond this
point, the quantity $R_p'$ is no longer given by $\beta^{-1}$ but we have
instead the behavior $R_p'\sim (\alpha/\Gamma_m)^{1/3}$ (the same happens
with $d$). A similar approach was adopted in Ref.~\cite{hk08b}, where $R_p'$
was estimated by using the approximation $\bar\Gamma\simeq\Gamma$ and
differentiating the volume weighted distribution $R^3\exp(-S)$. This gives
the equation $R_p'=3v_w/\beta(T_{R_p'})$, where $T_{R_p'}$ is the
temperature at which the bubbles of size $R_p'$ were nucleated. Using the
time-temperature relation to obtain $T_{R_p'}$ as a function of $R_p'$, this
equation can be solved numerically. For the gaussian approximation, we have
$\beta(T_{R_p'})\simeq 2\alpha^2(t_{R_p'}-t_m)$ and, using $R_p'\simeq
v_w(t-t_{R_p'})$, we obtain the analytic approximation (\ref{rpprigap}).

\section{Conclusions} \label{conclu}

We have studied the dynamics of bubble nucleation and growth in very strong
phase transitions of the universe. In particular, we have computed the
nucleation rate and the wall velocity, and we have tracked the progress of
the transition, using as a measure the fraction of volume which remains in
the false vacuum. We have identified several milestones in the evolution,
associated to the equality of the expansion and nucleation rates, the
nucleation of the first bubbles in a Hubble volume, the percolation of
bubbles, and specific values of the fraction of volume. We have studied the
number density of bubbles and the distribution of bubble sizes as functions
of time, and we have discussed the possible cosmic remnants.

For very strong phase transitions the wall velocity is $v_w\simeq 1$, and
the relevant quantities which determine the dynamics are the nucleation rate
$\Gamma$ and the Hubble rate $H$. The instanton action $S(T)$ which
determines the nucleation rate is a dimensionless quantity which does not
depend on the scale of the theory but only on the shape of the effective
potential. Using a simple yet physical model, we have considered model
parameters for which $S(T)$ has a minimum, in which  case the phase
transition can be very strong. As long as the minimum is not reached, the
main differences with weaker phase transitions are only quantitative. On the
other hand, if $S$ gets too close to its minimum, the dynamics becomes
qualitatively different. Besides, in this case the usual approximation of
linearizing $S$ around a time $t_*$ near the completion of the transition,
which leads to an exponentially growing nucleation rate, breaks down. In
particular, the parameter $\beta =d\log\Gamma/dt$, which characterizes the
duration of the phase transition and the typical bubble size,  will vanish
or become negative, which poses a problem for applying results obtained with
this approximation to physical models.

Nevertheless, around its minimum, $S$ is approximately quadratic, and the
nucleation rate can be approximated by a gaussian function of time. In this
case, the relevant dimensional parameter $\alpha$ is given by
$\alpha^2=({d^2\Gamma}/{dt^2})/(2\Gamma)$. We have solved analytically the
development of the phase transition for the gaussian approximation, both for
a static universe and for an exponentially growing scale factor. Ignoring
the scale factor is justified if the duration of the transition is short
enough, which in this case is determined by the time scale $\alpha^{-1}$.
For cases in which the duration is longer than this time, an exponential
scale factor is generally a good approximation. In the general case, the
time and length scales are determined by the parameter $\alpha$ as well as
by the value of the nucleation rate at the minimum, $\Gamma_m$. Thus, for
instance, the average bubble size at the end of the transition is $\bar
R\sim (\alpha/\Gamma_m)^{1/3}$. The gaussian approximation also allowed us
to determine analytically the limit in which the system remains stuck in the
metastable phase, which gives the bound $\Gamma_m/H^4\gtrsim\alpha/H$.

We have compared the analytic results for the gaussian and for the
exponential rates with those obtained with the fully numerical computation.
As we have seen, the exponential-rate approximation is valid for
$\beta\gg\alpha$. For $\beta\lesssim\alpha$ the gaussian approximation can
be used, and in the limit between the two regimes, namely, $\beta=\mathrm{a
\ few}\,\alpha$, both approximations are quite good.

For physical models, the parameter region for which the phase transition
occurs around the minimum of $S$ is relatively small. In our example model,
in which we varied a single parameter, only a 3\% of its range corresponds
to phase transitions of this kind. In any case, such strong phase
transitions are possible and may have important consequences, particularly
for the generation of gravitational waves. As we have shown, the dynamics of
a very strong phase transition cannot be obtained by direct extrapolation
from that of weaker phase transitions. In order to consider this possibility
in a simulation (and take into account the dynamics in a realistic way), a
gaussian nucleation rate should be considered instead of the usual
exponential approximation. In the case of gravitational waves, a possible
extrapolation for results which are functions of $\beta$ is to write $\beta$
in terms of a physical length such as the peak of the volume-weighted bubble
distribution. For a very strong phase transition, this length can be
obtained analytically from the gaussian approximation. In this case, the
extrapolation consists in the replacement $\beta\to
4.1(\Gamma_m/\alpha)^{1/3}$.

\section*{Acknowledgements}

This work was supported by FONCyT grant PICT 2013 No.~2786, CONICET grant
PIP 11220130100172, and Universidad Nacional de Mar del Plata, Argentina,
grant EXA793/16.

\end{document}